\documentclass[useAMS,usenatbib]{mn2e}

\usepackage{graphicx}
\usepackage{natbib}
\usepackage{subscript}
\usepackage{multicol}
\usepackage{amsmath,amssymb,latexsym} 
\usepackage{longtable}
\usepackage{aecompl}
\usepackage{lastpage}
\usepackage{pdflscape}
\usepackage{subfig} 
\usepackage{multirow}
\usepackage[T1]{fontenc}
\usepackage{lmodern}
\usepackage{aecompl}
\title[A dearth of OH/IR stars in the SMC]{A dearth of OH/IR stars in the Small Magellanic Cloud}

\author[S. R. Goldman et al. ]{Steven R. Goldman$^{1}$\thanks{E-mail: s.r.goldman@keele.ac.uk}, Jacco Th. van Loon$^{1}$, Jos\'e F. G\'omez$^{2}$, James A. Green$^{3,4}$, \newauthor Albert A. Zijlstra$^{5}$, Ambra Nanni$^{6}$, Hiroshi Imai$^{7}$, Patricia A. Whitelock$^{8,9}$, \newauthor Martin A. T. Groenewegen$^{10}$ and Joana M. Oliveira$^{1}$
\vspace{0.25cm} \\ 
$^{1}$Astrophysics Group, Lennard-Jones Laboratories, Keele University, ST5 5BG, UK\\ 
$^{2}$Instituto de Astrof\'{\i}sica de Andaluc\'{\i}a, CSIC, Glorieta de la Astronom\'{\i}a s/n, 18008 Granada, Spain \\
$^{3}$SKA Organisation, Jodrell Bank Observatory, Lower Withington, Macclesfield, Cheshire, SK11 9DL, UK \\
$^{4}$CSIRO Astronomy and Space Science, Australia Telescope National Facility, PO Box 76, Epping, NSW 1710, Australia\\
$^{5}$Jodrell Bank Centre for Astrophysics, Alan Turing Building, School of Physics and Astronomy, The University of Manchester,\\ Oxford Road, Manchester, M13 9PL, UK \\
$^{6}$Dipartimento di Fisica e Astronomia Galileo Galilei, vicolo dell'Osservatorio 3, 35122 Padova PD, Italy\\
$^{7}$Department of Physics and Astronomy, Kagoshima University, 1-21-35 Korimoto, Kagoshima 890-0065, Japan\\
$^{8}$South African Astronomical Observatory (SAAO), PO Box 9, 7935 Observatory, South Africa\\
$^{9}$Astronomy Department, University of Cape Town, 7701 Rondebosch, South Africa \\
$^{10}$Koninklijke Sterrenwacht van Belgi\"e,  Ringlaan 3, B-1180 Brussels, Belgium }

\date{Accepted 2017 XXXXXXX XX. Received 2016   XXXXXXX XX}
\pubyear{2017}

\begin{document}
\pagerange{\pageref{firstpage}--\pageref{lastpage}}
\maketitle
\label{firstpage}

\begin{abstract}
\noindent 
We present the results of targeted observations and a survey of 1612-, 1665-, and 1667-MHz circumstellar OH maser emission from asymptotic giant branch (AGB) stars and red supergiants (RSGs) in the Small Magellanic Cloud (SMC), using the Parkes and Australia Telescope Compact Array radio telescopes. No clear OH maser emission has been detected in any of our observations targeting luminous, long-period, large-amplitude variable stars, which have been confirmed spectroscopically and photometrically to be mid- to late-M spectral type. These observations have probed $3-4$ times deeper than any OH maser survey in the SMC. Using a bootstrapping method with LMC and Galactic OH/IR star samples and our SMC observation upper limits, we have calculated the likelihood of not detecting maser emission in any of the two sources considered to be the top maser candidates to be less than 0.05\%, assuming a similar pumping mechanism as the LMC and Galactic OH/IR sources. We have performed a population comparison of the Magellanic Clouds and used \textit{Spitzer} IRAC and MIPS photometry to confirm that we have observed all high luminosity SMC sources that are expected to exhibit maser emission. We suspect that, compared to the OH/IR stars in the Galaxy and LMC, the reduction in metallicity may curtail the dusty wind phase at the end of the evolution of the most massive cool stars. We also suspect that the conditions in the circumstellar envelope change beyond a simple scaling of abundances and wind speed with metallicity.

\end{abstract}


\section{Introduction}

The Magellanic Clouds have considerably different Asymptotic Giant Branch (AGB) populations resulting from different star formation histories (SFHs). The Small Magellanic Cloud (SMC), with a stellar mass of $3.7 \times 10^8$ M$_{\odot}$ \citep{2012ApJ...761...42S}, is nearly a quarter of the stellar mass of the Large Magellanic Cloud (LMC) at $1.7 \times 10^9$ M$_{\odot}$ \citep{2004AJ....127.1531H,2009AJ....138.1243H}. Within the LMC, a peak in star formation around 100 Myr ago \citep{2009AJ....138.1243H} has resulted in favourable conditions for bright maser emitting evolved stars. Moreover, according to the recent work by \citet{2017MNRAS.465.4817S} based on the stellar evolutionary tracks computed with the \textsc{parsec} \citep{2012MNRAS.427..127B} and \textsc{colibri} \citep{2017ApJ...835...77M} codes, the minimum mass of a Hot-Bottom Burning (HBB) star, which provides an oxygen-rich environment favorable for maser-emission, is around 4.2 M$_{\odot}$ for Z=0.006; the exact value of the mass will depend on the metallicity and on the underlying stellar model. As a consequence, given the 100 Myr evolution time of $5-6$ M$_{\odot}$ stars, this has resulted in a disproportionally large population of sources undergoing HBB \citep{2015MNRAS.454.4235D,2017MNRAS.465.4817S}. Conversely, the SFH of the SMC derived by \citet{2004AJ....127.1531H} shows a minimum in the star formation around 100 Myr ago, which has likely resulted in a smaller population of massive oxygen-rich AGB stars within the SMC. Even if the minimum in the star formation rate (SFR) is not confirmed by the most recent SFH derived by \citet{2015MNRAS.449..639R}, the SFR by \citet{2004AJ....127.1531H} is in good agreement with the one derived by \citet{2015MNRAS.449..639R} about 100 Myr ago. The more recent SFR is difficult to constrain using Hubble Space Telescope data as it requires a wide survey area \citep{2013MNRAS.431..364W}. 

AGB stars of masses 1 M\textsubscript{$\odot$} $\lesssim$ M $\lesssim 8$ M\textsubscript{$\odot$}, and some red supergiants (RSGs) with masses M $\gtrsim$ 8 M\textsubscript{$\odot$}, go through periods of intense mass-loss. Near the end of their lifetimes, AGB stars will lose $50 - 80\%$ of their initial mass, at a rate of up to $10^{-4}$ M$_{\odot}$ yr$^{-1}$ \citep{1999A&A...351..559V}, and contribute a significant amount of chemically enriched material to the interstellar medium. They may also be the largest contributors of dust to the universe \citep{1989IAUS..135..445G}, but this remains controversial as the effects of metallicity and luminosity on the dust contribution from these stars are still quantitatively unclear, as are the contributions from other major sources, novae and supernovae. The mass loss has an important effect on the evolution of the star itself truncating its evolution as a cool (super)giant and, in the case of red supergiants, affecting the conditions for the unavoidable core-collapse supernova.

Towards the end of their lives, AGB and RSG stars will become obscured by circumstellar dust. This dust allows the stars to lose mass in the form of a stellar ``superwind'' \citep{Iben1983}. This wind is driven by radiation pressure on the dust grains that form at several stellar radii, though how the outflow is launched from the stellar photosphere is still not clear. As the dust grains move outward, they drag along gas and create winds that reach up to 30 km s$^{-1}$. The dust is composed of either carbonaceous or silicate dust grains depending on the ratio of carbon-to-oxygen within the stellar envelope. The chemical composition of the dust and envelope can be significantly altered by internal processes like HBB and/or third dredge-up events.

Evolved oxygen-rich AGB stars are known for exhibiting OH maser emission at 1612 MHz. While carbon-rich evolved AGB or ``carbon stars'' tend to be the more-obscured than their oxygen-rich counterparts, maser emission does not occur within them and thus we will exclude them from this study. Evolved AGB stars are significantly obscured by dust, which leaves them opaque at visible wavelengths. The stellar light from these stars is re-radiated by the dust grains making them bright in the infrared (IR). From these unique characteristics these stars have been designated as OH/IR stars. OH/IR stars are generally brightest in their OH maser emission at 1612 MHz, but also exhibit H\textsubscript{2}O and SiO masers. The 1612-MHz maser transition occurs at several hundred stellar radii where the outer-most H\textsubscript{2}O is photodissociated by interstellar ultraviolet radiation. As the transition probes the final outward wind speeds propagating towards and away from us, we observe a double-horned maser profile that is created from the Doppler-shifted 1612-MHz transition. By taking half the peak separation of the maser profile, we get the expansion velocity of the envelope of the star. This expansion velocity can be used to test and refine dust-driven wind theory. The expansion velocity has been shown to follow a relationship of: $v$\textsubscript{exp} $\propto \psi^{1/2}L^{1/4}$, where $L$ is the luminosity, and $\psi$ is the dust-to-gas ratio \citep{2017MNRAS.465..403G}. The dust-to-gas ratio has shown strong empirical evidence to scale approximately with the metal content \citep{1994A&A...286..523H,2000A&A...354..125V,Elitzur2001,2012arXiv1210.0983V} and thus we expect the expansion velocity should show a dependence on both metallicity and luminosity, a phenomenon that has been seen observationally in the LMC and Galactic samples \citep{1992ApJ...397..552W,2004MNRAS.355.1348M,2017MNRAS.465..403G}.

\subsection{Previous searches for oxygen-rich dusty stars in the SMC}

Two OH maser studies have targeted SMC circumstellar OH maser emission in the past (discussed further in Appendix A). The deepest observation resulted in a noise level $\sim$8 mJy. The Infrared Astronomical Satellite \citep[\textit{IRAS};][]{1984ApJ...278L...1N}, the Midcourse Space Experiment \citep[\textit{MSX};][]{2001AJ....121.2819P}, the Two-Micron All-Sky Survey (\textit{2MASS}) and more recently the \textit{Spitzer Space Telescope} \citep{2004ApJS..154....1W} have revealed large samples of AGB and RSG stars in the SMC \citep{1992ApJ...397..552W,2011AJ....142..103B,2017ApJ...834..185K}. However the search for deeply embedded sources continues. Based on MSX and 2MASS data, the IR-bright sample from the MSX survey was expected to be composed of one-third carbon stars and one-half oxygen-rich stars in the SMC. Instead, two-thirds of the sample were found to be carbon stars, and no heavily obscured oxygen-rich stars were detected, with the remaining third composed of RSGs, post-AGB stars, YSOs, PN, and oxygen-rich stars without significant dust. 

Other recent searches have discovered a few candidate OH/IR stars, some of which have been observed in the IR \citep{2015ApJ...811..145J,2017ApJ...834..185K}, and some at radio frequency, but only a few deeply embedded evolved stars like those within the LMC or our galaxy, either carbon- or oxygen-rich, have been found in the SMC \citep{2005A&A...442..597V,vanLoon:2008iv,2007MNRAS.376.1270L}. It is not a requirement for OH/IR stars to have a high self-extinction, but as the maser luminosity is dependent on the dust, the most luminous maser sources are expected to.

As was noted by \citet{2016MNRAS.457.1456V}, the most obscured carbon stars in the LMC and SMC have initial masses of $2.5-3 M_{\odot}$ and $\sim1.5 M_{\odot}$, respectively, as a result of different SFHs. This has lead to a difference in initial stellar masses in their Thermally-Pulsing AGB (TPAGB) phase in the Magellanic Clouds.

Moving to the higher mass stars, previous searches have found a small sample of dusty RSGs in the SMC, but they are not obscured to the same degree as the OH maser-emitting LMC RSGs \citep{2010AJ....140..416B}. They are both warmer and less dusty than the massive O-rich OH maser-emitting AGB stars and RSGs (called OH/IR stars) in the LMC and the Galaxy. While these stars are few, the results of the Surveying the Agents of Galaxy Evolution in the tidally stripped, low metallicity Small Magellanic Cloud (SAGE-SMC) survey \citep{2011AJ....142..102G} suggest that the small known sample of dusty RSGs within the Magellanic Clouds contributes more dust than the entire sample of optically bright RSGs \citep{2010AJ....140..416B}. Hence it is critically important that we obtain measurements of the expansion velocities of these sources to constrain current wind driving models. 

\begin{table*}
\centering
\caption[The SMC maser target sample characteristics]{The most promising maser candidates in the SMC. Listed values are derived bolometric luminosities from SED fitting by \textsc{dusty} models (described further in Section 4.5), $I$-band pulsation periods and amplitudes from \citet{2011AcA....61..217S}, $J-K$ colours from 2MASS \citep{2003yCat.2246....0C}, \textit{Spitzer} MIPS \citep{2004ApJS..154...25R} 24 $\mu$m fluxes ($F_{24}$) from \citet{2011AJ....142..103B}, and spectral types (van Loon et al. 2008, and references therein).} 
\centering
\begin{tabular}{ l c c c c c c}
\hline 
Object & 
$L$ & 
Pulsation period &
Pulsation amplitude &
$J-K$ &
$F_{24\mu m}$ &
Spectral 
\\
name & 
($10^{3}$ L$_{\odot}$) &
(d) & 
(mag) &
(mag) &
(mJy) &
type 
 \\
\hline 
\textit{Parkes targets} \\ 
IRAS\,00483$-$7347 &
\llap{13}8 &
\llap{1}859&
1.7 &
2.8 &
\llap{6}60.0 &
M8 
 \\
IRAS\,F00486$-$7308 &
\llap{4}1 & 
\llap{1}062&
2.7 &
1.6 &
91.1 &
M4 
 \\
IRAS\,00591$-$7307 &
\llap{7}8 & 
\llap{1}092 &
1.9 &
1.3 &
95.4 &
M5\rlap{e} 
 \\
IRAS\,01074$-$7140 &
\llap{3}4 &
523&
2.8 &
1.2 &
396 &
M5\rlap{$^d$}
\\
\\
\textit{ATCA sources}\\
OGLE SMC-LPV-15504 & 
\llap{4}5 & 
543 & 
0.8&
1.3 &
31.6 &
M3$-$4\rlap{$^c$}  
\\
2MASS J01033142$-$7214072 & 
7 &
471&
1.8 &
3.7 &
19.2 &

\\
HV 12149 & 
\llap{4}8 & 
769 & 
2.3 &
1.4 &
12.7 &
M8\rlap{$^b$} 
\\
OGLE SMC-LPV-14322 &
\llap{1}1 &  
483 & 
1.9 &
3.5 &
19.0 &

\\
2MASS J00592646$-$7223417 & 
8 & 
&
&
\llap{$>$}4.4 & 
18.9 &

\\
BMB-B75 &
\llap{5}8 & 
761 &
2.1 &
1.3 &
92.0 &
M6\rlap{$^a$} 
\\
OGLE J004942.72$-$730220.4 & 
8 &
563&
0.4 &
3.9 &
27.8 &

\\
MSX SMC 018 &
\llap{6}0 &
897 &
2.5&
2.5 &
\llap{2}62.0 &
 \\ \hline
\end{tabular}
\flushleft{\textit{References}:
$^a$\citet{1980ApJ...242..938B}
$^b$\citet{2015A&A...578A...3G}
$^c$\citet{2007ApJ...660..301M}
$^d$\citet{1989MNRAS.238..769W}.}
\label{smc_maser_candidates} 
\end{table*}

\section{Description of the Sample}

Past observations have succeeded in detecting interstellar OH maser emission in the SMC \citep{2013MNRAS.432.1382B}, but have failed to detect circumstellar OH maser emission \citep{2012arXiv1210.0983V}. We have targeted luminous evolved stars to investigate and scrutinize the most promising candidates for circumstellar maser emission to date (Table \ref{smc_maser_candidates}). Our targets were chosen based on results from \citet{1980ApJ...242L..13E,1985ApJS...57...91E,1989MNRAS.238..769W,1998A&A...332...25G}, as well as pulsation periods newly derived from the Optical Gravitational Lensing Experiment (OGLE) \citep{2011AcA....61..217S}.

\begin{figure}
 \centering
 \includegraphics[width=1\columnwidth]{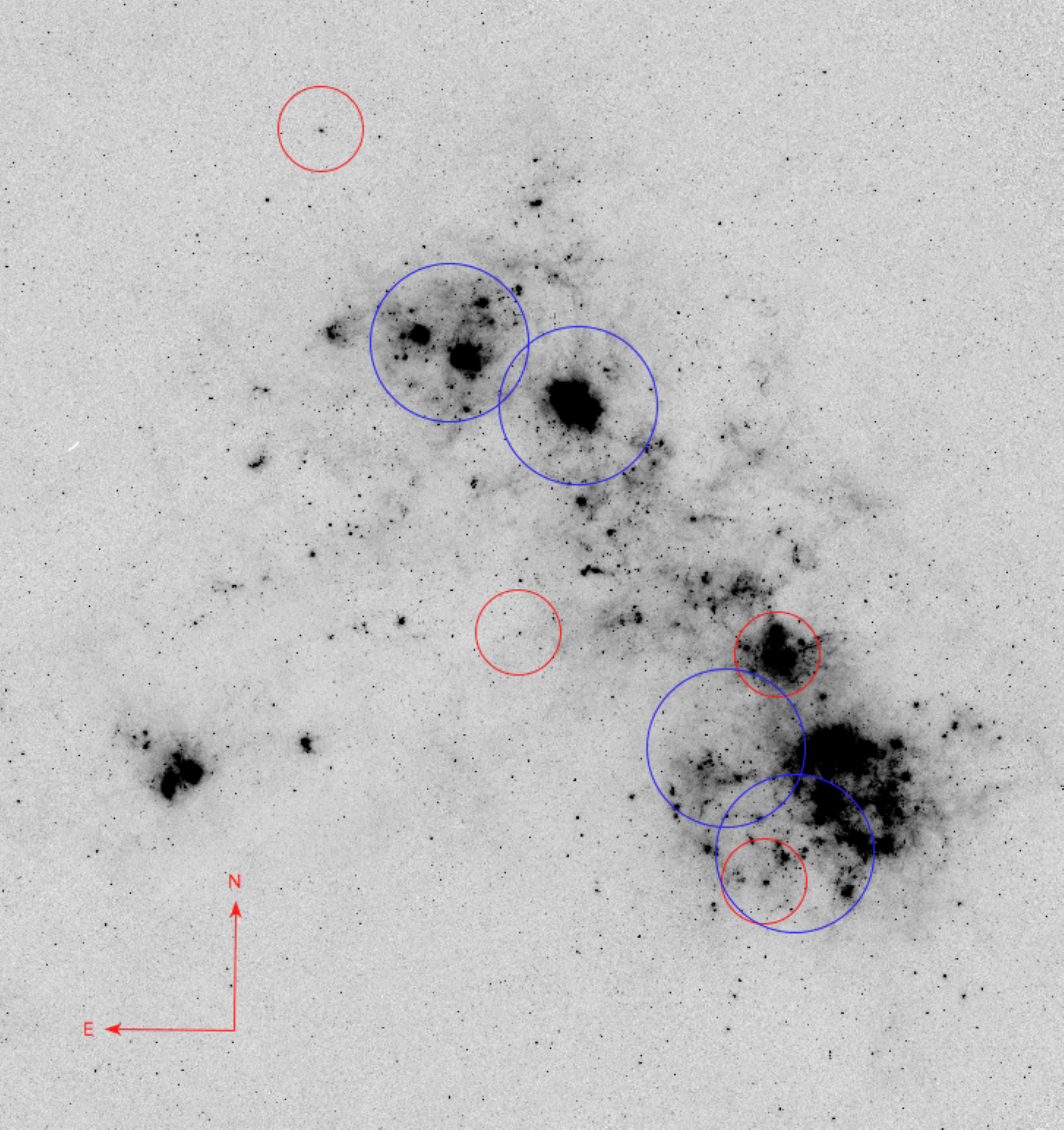} 
 \caption[SMC observation fields]{The \textit{Spitzer} MIPS 24 $\mu$m image of the SMC overlaid with the full width at half-maximum (FWHM) of the primary beam at 1612 MHz of our ATCA interferometric observations of 29 arcminutes (shown in blue) and those of the Parkes single-dish observations at 14 arcminutes (shown in red).}
 \label{smc_fields}
\end{figure}

Our observations include four sources targeted for being dusty evolved oxygen-rich sources with the Parkes radio telescope as well as an ATCA survey with four fields spread across the SMC. With Parkes we have targeted four of the best candidates for OH maser emission in the SMC. With the ATCA's FWHM primary beam at 1612 MHz of 29 arcminutes, we have surveyed a significant portion of the SMC bar region (Fig. \ref{smc_fields}). We have obtained spectra for seven sources that are known to be mid$-$late M-type oxygen-rich sources, one source that is likely to be oxygen-rich (MSX SMC 018), and four sources that are likely carbon stars. 

We expect to detect maser emission in several of our observed sources. The 1612-MHz maser is pumped by IR radiation at 35 $\mu$m with a typical efficiency of 23\% (discussed further in Section 4.4). Simplistically assuming this efficiency, the maser flux of our targeted source IRAS 00483$-$7347 should be over 19 times the noise level of the observation. 

\subsection{Observations}

Radial velocities mentioned in this paper are in a barycentric reference frame. We adopt a systemic radial velocity of $\approx 146$ km s$^{-1}$ for the SMC but radial velocities of stars and gas within the SMC vary by about $\pm 50$ km s$^{-1}$ around this value, stemming from a combination of dispersion and gradient on the sky \citep{1997A&AS..122..507H,1999MNRAS.302..417S}.

\subsubsection{Parkes observations}

We present archival (reprocessed) Parkes observations taken in 2003 from August 18 to 20 \citep{2004MNRAS.355.1348M}, and new observations taken in 2005 from July 7 to 13 and August 13 to 19, observing the 1612-MHz OH satellite line. With the multibeam receiver and multibeam correlator, we used a dual polarization setup with 8-MHz bandwidth (1489 km s$^{-1}$) and 8192 channels centered at 117 km s$^{-1}$ yielding a 0.18 km s\textsuperscript{$-1$} channel\textsuperscript{$-1$} velocity resolution; the observations used a frequency-switching calibration technique with a switching frequency of 0.5 MHz (93 km s$^{-1}$ at 1612 MHz) and a 14 arcminute beam. IRAS 00591$-$7307 was observed in 2003, IRAS F00486$-$7308 and IRAS 01074$-$7140 in 2005, and IRAS 00483$-$7347 in both epochs.

\begin{table*}
\centering
\caption[SMC observation fields]{Results of OH maser searches in the Small Magellanic Cloud. Values for the rms are taken from displayed spectra for Parkes, and calculated over the region of the three-dimensional datacube of which the spectrum was extracted for ATCA.}
\begin{tabular}{lcccc}
\hline &
\multirow{2}{*}{RA} &
\multirow{2}{*}{DEC}& 
On-source & 
\multirow{2}{*}{rms}
\\
Target name & 
\multirow{2}{*}{(J2000)} & 
\multirow{2}{*}{(J2000)} & 
Integration &
\multirow{2}{*}{(mJy)}  \\
 & 
& 
& 
time (h) & 
 \\
\hline 
\textit{Parkes observations}\\ 
IRAS\,00483$-$7347 &
00h 50m 09s  &
$-73^{\circ} \, 31^{\prime} \, 29^{\prime \prime}$ &
58.8 &
3.7 
 \\
IRAS\,F00486$-$7308 &
00h 50m 31s  &
$-72^{\circ} \, 51^{\prime} \, 30^{\prime \prime}$ &
31.1 &
4.5
 \\
IRAS\,00591$-$7307 &
01h 00m 48s  &
$-72^{\circ} \, 51^{\prime} \, 02^{\prime \prime}$ &
10.0 &
7.7 
 \\
IRAS\,01074$-$7140 &
01h 09m 02s &
$-71^{\circ} \, 24^{\prime} \, 10^{\prime \prime}$ &
61.9 &
3.4
\\ \\
\textit{ATCA observations}\\ 
SMC North  & 01h 03m 60s    & $-72^{\circ} \, 01^{\prime} \, 00^{\prime \prime}$ &  \llap{1}1.4 & 8      \\
SMC Centre & 00h 59m 00s    & $-72^{\circ} \, 10^{\prime} \, 60^{\prime \prime}$ &  \llap{1}0.0 & 9       \\
BMB-B75   & 00h 52m 13s    & $-73^{\circ} \, 08^{\prime} \, 53^{\prime \prime}$ &  \llap{1}1.2 & \llap{1}0       \\
SMC South  & 00h 48m 60s    & $-73^{\circ} \, 25^{\prime} \, 60^{\prime \prime}$ &  8.4 & 9 \\
\hline
\label{smc_obs}    
\end{tabular}
\end{table*}

\subsubsection{ATCA observations}
The data were taken in 2012 between January 1 and January 5, and in 2015 on June 12. Four observations of varying integration times between 8 and 11 hours were done for each targeted region (displayed in Table \ref{smc_obs}). The observations were done with the CFB 1M-0.5k correlator configuration with zoom-bands \citep{2011MNRAS.416..832W} and the 6A array configuration, and observed all four OH maser transitions at 18 cm (1612, 1665, 1667, and 1720 MHz). The bandwidth used was 2.5 MHz with 5121 channels, corresponding to zoom-band velocity widths of $\sim$ 465 km s$^{-1}$ and a velocity resolution of 0.09 km s$^{-1}$ centred at $\sim50$ km s$^{-1}$. The FWHM of the primary beam of ATCA at 1612 MHz is 29 arcminutes. Three of the observations were blind maser searches that targeted two star formation regions in the north and another in the south (Fig. \ref{smc_fields}). These observations used PKS 1934-638 and PKS 0823-500 as flux and bandpass calibrators and PKS 2353-686 and PKS 0230-790 as secondary calibrators. The fourth observation targeted the region covering the evolved star BMB-B75 toward the southern SMC field and used PKS 1934-638 as a bandpass and a flux calibrator and PKS J0047-7530 as a secondary calibrator. 

\subsection{Data reduction}

\subsubsection{Parkes data reduction}
Using the \textsc{casa asap} toolkit \citep{2007ASPC..376..127M}, spectra were extracted for each of our four targeted sources. Employing a frequency-switching calibration, out-of-band scans were subtracted to ensure a flat baseline. The spectra of IRAS 00483$-$7347 synthesised from a combination of both epochs and those of the remaining targets are displayed and discussed in  Section 3. The resulting spectra were resampled to a resolution of 0.5 km s$^{-1}$. This was done to mitigate large spikes from RFI while preserving adequate spectral resolution.

\subsubsection{ATCA data reduction}
The ATCA data were inspected, flagged, and calibrated using \textsc{miriad} \citep{1995ASPC...77..433S}. The visibility data were then transformed into three-dimensional data cubes, weighting the visibilities naturally. The source-finding package \textsc{duchamp} \citep{2012MNRAS.421.3242W} was used to search for maser sources within the FWHM of each of the fields. For the resulting peaks, our intended targets, and any other potential targets from \citet{2011AJ....142..103B}, \citet{2012ApJ...753...71R}, or SIMBAD, a spectrum was extracted for a region the size of the synthesized beam ($\sim 7^{\prime \prime}$). All overlapping fields were mosaicked, and \textsc{duchamp} was used to search within the ATCA's FWHM primary beam, centred between the mosaicked fields. The resulting spectra were resampled to a resolution of 0.5 km s$^{-1}$, as was done with the Parkes spectra. 

\section{Results}
\begin{figure*}
 \centering
 \includegraphics[width=8.7cm]{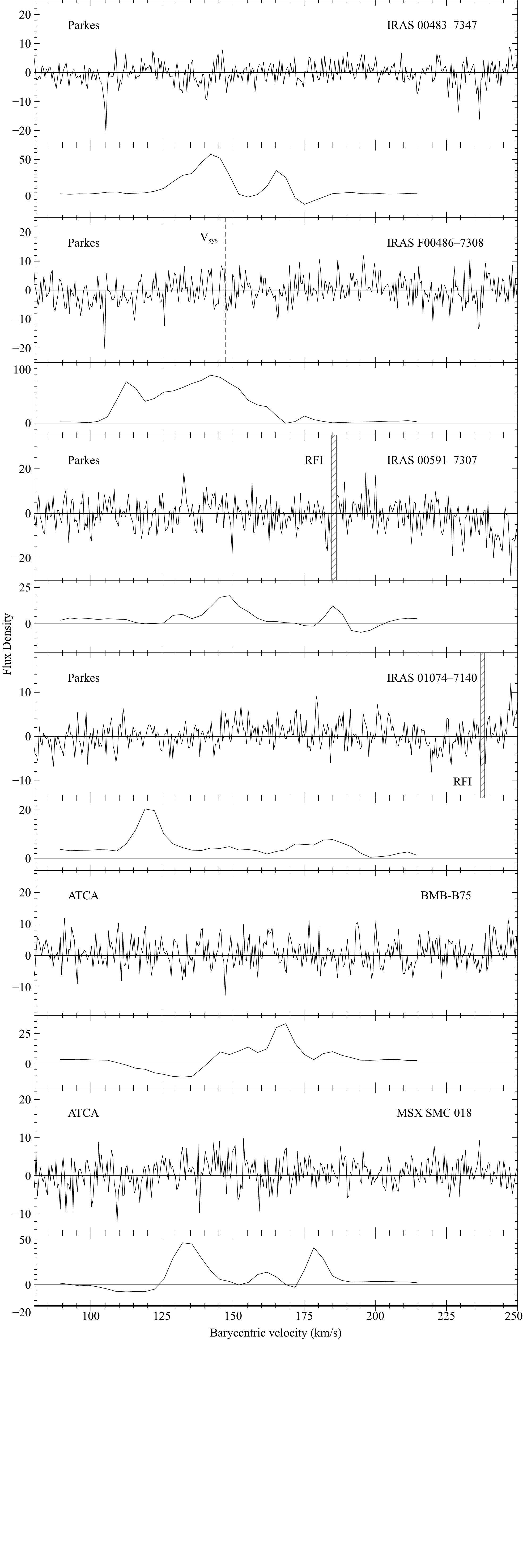} \hspace{-0.1cm}
 \includegraphics[width=8.7cm]{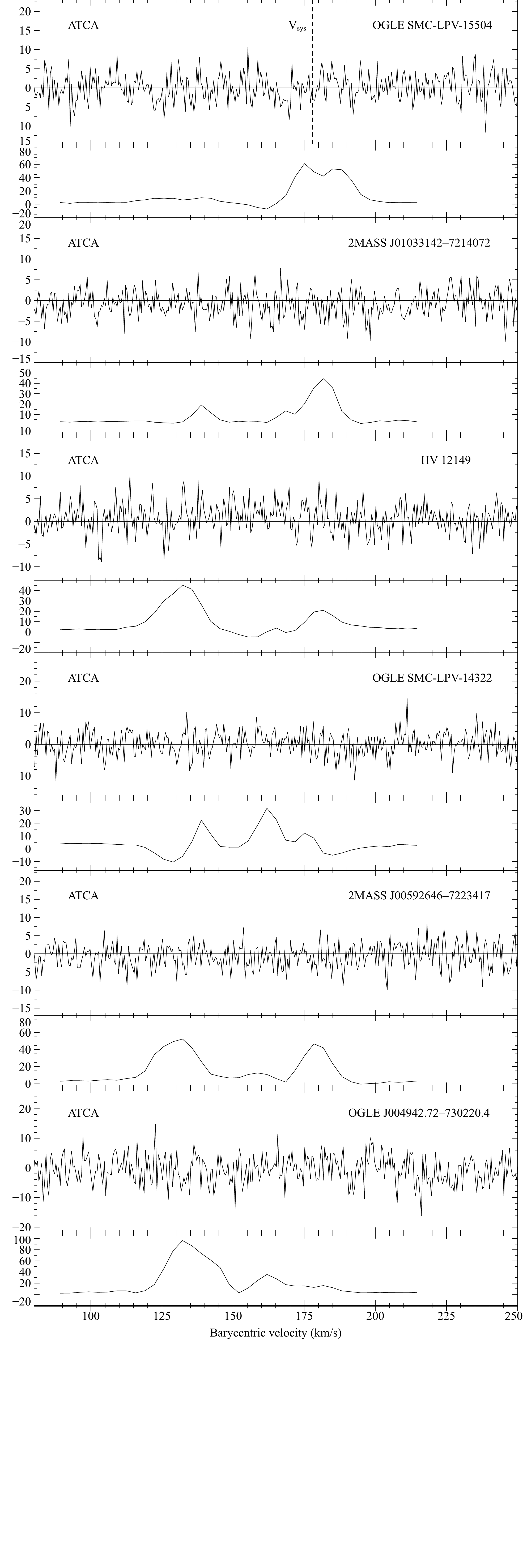} \vspace{-4.1cm}
 \caption{OH 1612-MHz maser observations of our SMC stars resampled to a velocity resolution of 0.5 km s$^{-1}$. Below our maser spectra (flux density in mJy) are HI spectra (flux density in Jy) to indicate the likely systemic velocity of the source. Also shown are measured systemic velocities (V\textsubscript{sys}) for IRAS F00486$-$7308 and OGLE SMC-LPV-15504 from \citet{2015A&A...578A...3G} and \citet{2003AJ....126.2867M}, respectively; areas of Radio Frequency Interference (RFI) are also indicated.}
 \label{smc_masers}
\end{figure*}

Our targeted and blind maser searches within the SMC have yielded no clear OH maser emission. This is the case for the 1612-MHz OH observations (Fig. \ref{smc_masers}), as well as the 1665-, 1667-, and 1720-MHz OH maser observations; the 1665-MHz OH mainline observations are shown in Appendix C. The average sensitivities of the 1612-MHz observations are listed in Table \ref{smc_obs}. While seemingly unsuccessful, these observations have in fact given us valuable upper limits. In the following sections we will employ our new limits to draw conclusions about the absence of an OH/IR population in the SMC, and what we can predict for future maser searches. 

The H\,{\sc i} column density has shown to be a good indicator of the likely systemic velocities of our sources. In addition to the Parkes and ATCA observations, we have also included 1420-MHz H\,{\sc i} data from the combined ATCA and Parkes multibeam H\,{\sc i} maps \citep{1999MNRAS.302..417S}. The survey covers an area of 20 square degrees with a velocity resolution of 1.6 km s$^{-1}$, a velocity range from $\sim 90 - 215$ km s$^{-1}$, and an angular resolution of 98 arcseconds (FWHM). For each source in our SMC sample, a spectrum was extracted using a 1-arcminute region centred on the sources. The resulting spectra have been plotted below each maser spectra. While some of our sources lie outside these H\,{\sc i} regions, they provide supporting evidence for sources which lie in higher density regions of the SMC. Within the observations we find no indication of maser emission.

 
\begin{table*}
\centering
\caption[The upper limits for circumstellar OH maser emission in the SMC]{The flux density upper limits for circumstellar OH maser emission for our SMC sample. Listed values are \textit{Spitzer} MIPS 24 $\mu$m flux ($F_{24}$) from \citet{2011AJ....142..103B}, predicted 35 $\mu$m flux values taken from the best-fit \textsc{dusty} model of the source (explained in Section 4.5), predicted OH maser flux ($F_{\textrm{OH,\,predicted}}$) assuming our median masing efficiency of 23$\%$ $F_{35}$ photons to $F_{\textrm{OH}}$ photons, and the noise level ({$\sigma$}) of the observation (both combined observations for IRAS 00483$-$7347. Also shown are the upper limits for maser efficiency (maximum efficiency) and the probability of a $3 \sigma$-level detection for each targeted observation calculated with our bootstrapping method using the maser efficiencies of the LMC and Galactic samples from \citet{2017MNRAS.465..403G}. } 
\centering
\begin{tabular}{lcccccc}
\hline
Object &
$F_{24\mu \textrm{m}}$ &
$F_{35\mu \textrm{m}, \textrm{predicted}}$ &
$F_{\textrm{OH,\,predicted}}$&
$\sigma$ &
Maximum&
Likelihood
 \\
name &  
(mJy) &
(mJy) &
(mJy) &
(mJy) &
efficiency &
of detection 
\\
\hline
\textit{Parkes targets}\\ 
IRAS\,00483$-$7347 &
\llap{6}60.0 &
\llap{30}7 &
\llap{7}1 &
3.7 &
3.6\% &
\llap{9}7.7\% \\
IRAS\,F00486$-$7308 &
91.1 &
\llap{3}9 &
9 &
4.5 &
\llap{3}4.5\% &
\llap{4}0.7\% \\
IRAS\,00591$-$7307 &
95.4 &
\llap{4}4 &
\llap{1}0 &
7.7 &
\llap{5}2.1\% &
\llap{2}3.3\% \\
IRAS\,01074$-$7140 &
\llap{3}96.0 &
\llap{23}5 &
\llap{5}4 &
3.4 &
4.3\% &
\llap{9}7.7\% \\
\\
\textit{ATCA sources}\\ 
OGLE SMC-LPV-15504 & 
31.6 &
\llap{1}3 &
3 &
7.1 &
\llap{15}8.9\% &
5.8\% \\
2MASS J01033142$-$7214072 &
19.2 &
8 &
1 &
6.3 &
&
\\
HV 12149 & 
12.7 &
7 &
2 &
6.5 &
\llap{27}8.0\% &
1.2\% \\
OGLE SMC-LPV-14322 &
19.0 &
\llap{1}1 &
3 &
7.6 &
&
\\
2MASS J00592646$-$7223417 &
18.9 &
\llap{1}3 &
3 &
6.5 &
&
\\
BMB-B75 &
92.0 &
\llap{4}3 &
\llap{1}0 &
9.5 &
\llap{6}6.5\% &
\llap{1}9.7\% \\
OGLE J004942.72$-$730220.4 &
27.8 &
\llap{1}4 &
3 &
\llap{1}0.0 &
&
\\
MSX SMC 018 &
\llap{2}62.0 &
\llap{11}6 &
\llap{2}7 &
7.2 &
\llap{1}8.7\% &
\llap{6}7.5\% \\
\hline
\end{tabular}
\label{limits} 
\end{table*}

\subsubsection*{Parkes targets}

Our four Parkes targets are IRAS 00483$-$7347, IRAS F00486$-$7308, IRAS 00591$-$7307, and IRAS 01074$-$7140. All of these sources have infrared characteristics typical of OH maser emitting AGB stars and RSGs (Table 1) and show 10 $\mu$m silicate features in emission, confirming they are oxygen-rich \citep{1995ApJ...449L.119G,2011ApJS..196....8L}. IRAS 00483$-$7347 is our best candidate for OH maser emission with an expected mass of $6-7$ M$_{\odot}$ \citep{2015MNRAS.454.4235D}, and has been found to be the most massive HBB AGB star in the SMC from its large Rb enhancement \citep{2009ApJ...705L..31G}. It has been suggested to be a strong candidate for a super-AGB star \citep{2009A&A...506.1277G}, which should not affect the maser emission, but may explain its peculiarity among the other SMC stars. IRAS F00486$-$7308 has shown strong molecular bands around 3 $\mu$m, indicative of an advanced evolutionary stage where the star is cooler with stronger pulsations capable of extending the molecular atmosphere \citep{vanLoon:2008iv}. IRAS 01074$-$7140 is very bright in the IR ($F_{24}$ = 0.4 Jy) but does not have a particularly reddened colour or long pulsation period. This source has also shown to be Li-enhanced, indicative of HBB \citep{1989MNRAS.238..769W,1995ApJ...441..735S}.

\subsubsection*{ATCA sources}

With our ATCA survey of the SMC we have obtained spectra for eight sources that we have considered good maser candidates. Four of these sources (OGLE SMC-LPV-15504, HV 12149, BMB-B75, and MSX SMC 018) have shown 10 $\mu$m silicate features in emission. OGLE SMC-LPV-15504 has a known systemic velocity of $176.8\pm0.3$ km s$^{-1}$ \citep{2003AJ....126.2867M} from spectroscopic observations of the Ca {\sc ii} triplet, which greatly simplifies our search in velocity space. HV 12149 has an initial mass of around $4-5$ M$_{\odot}$, and is likely to have just initiated HBB \citep{2015A&A...578A...3G,2015MNRAS.454.4235D,2017MNRAS.465.4817S}. This source is unique with its late spectral type yet bright $K$-band flux of $M_{K}$= 8.6 mag \citep{2003yCat.2246....0C}, typical of a RSG. After observing BMB-B75, \citet{2015ApJ...811..145J} found that the far-IR emission detected by \textit{Herschel} was due to a spatially coincident galaxy. The mid-IR IRS spectrum of the source shows red-shifted (z $\sim$ $0.16$) emission peaks of Polycyclic Aromatic Hydrocarbons (PAHs) at 6.3 and 11.3 $\mu$m and Ne \textsc{iii} at 12.8 and 15.5 $\mu$m from the background galaxy \citep{2011ApJS..196....8L,2017ApJ...834..185K}, but the silicate emission detected from the source appears unaffected. This source has also been found to have 6 cm continuum \citep{2012SerAJ.184...93W}, emission likely emanating from the background galaxy. MSX SMC 018 was categorised as an M-type star by \citet{vanLoon:2008iv} who used 3$-$4 $\mu$m spectra for classification, after which a spectral type of M7 was determined by \citet{2009A&A...506.1277G}. The source shows deep TiO bands, characteristic of oxygen-rich evolved AGB (as opposed to RSG) stars, and weak crystalline silicate features \citep{2017ApJ...834..185K}. The remaining four sources, 2MASS J01033142$-$7214072, OGLE SMC-LPV-14322, 2MASS J00592646$-$7223417, and OGLE J004942.72$-$730220.4, are evolved AGB stars within our observed fields that (as we will discuss in Section 4.5) are likely to be carbon stars.

\section{Discussion}

\subsection{A dearth of OH masers in the SMC}

\begin{figure}
 \centering
 \includegraphics[width=\columnwidth]{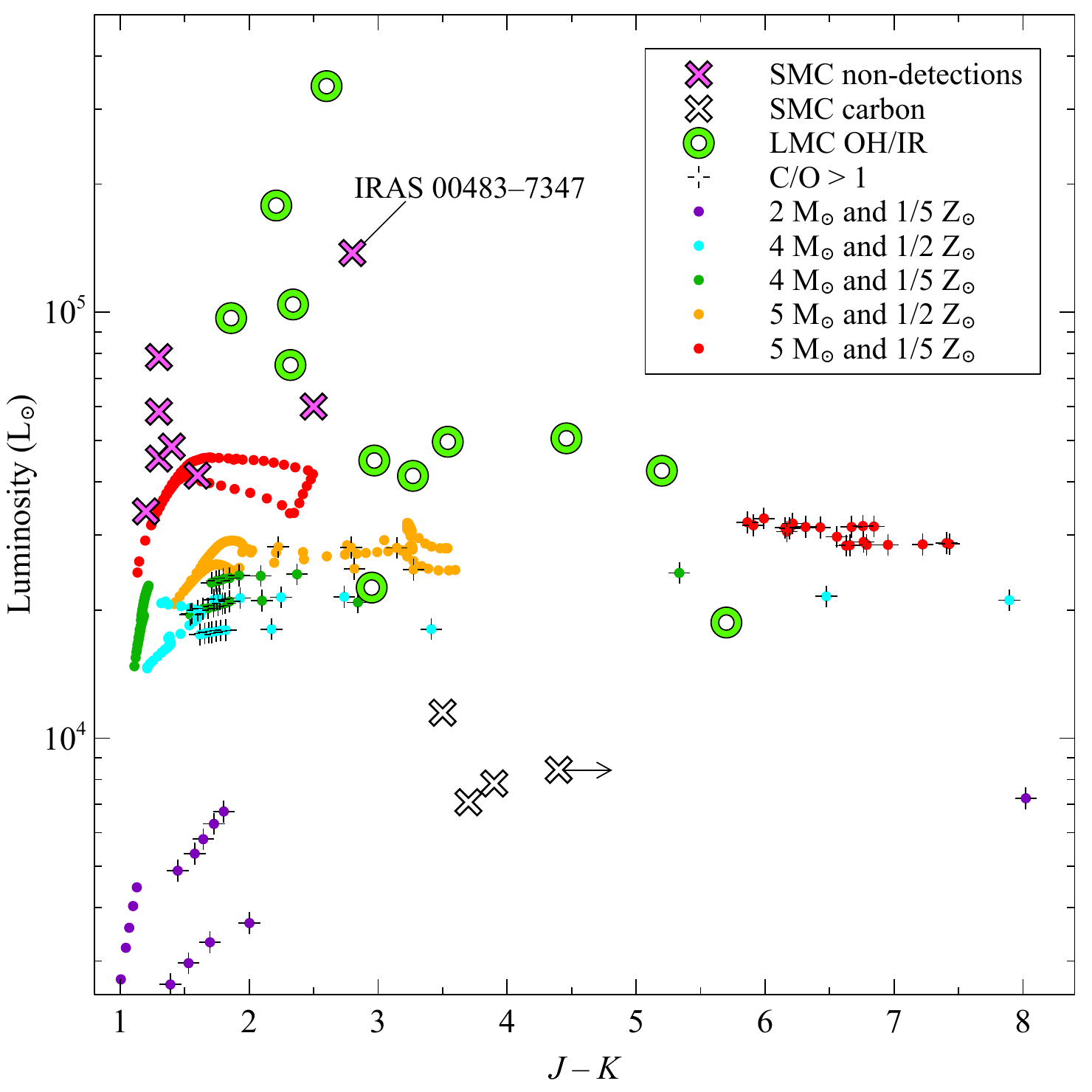}
 \caption[The luminosity as a function of $J-K$ colour]{The luminosity as a function of $J-K$ colour for our samples in the SMC and LMC. Also shown are \textsc{parsec-colibri} stellar isochrones \citep{2017ApJ...835...77M} corresponding to 2, 4, and 5 solar mass stars in their TPAGB phase for both the LMC and SMC metallicities; also marked on the isochrones are the points at which the carbon-to-oxygen ratio is greater than one.}
 \label{smc_L_vs_J-K}
\end{figure}

In trying to understand the lack of maser emission in any of our SMC candidates we will first look at our most promising SMC sources for maser emission, IRAS 00483$-$7347. This observation had a 59-hour integration and targets our most luminous candidate. It is possible that the source is uniquely void of maser emission. However, assuming the source is a maser-emitting source, we should be able to detect maser emission with a $F_{35}$-to-F\textsubscript{OH} pumping efficiency down to 3.6\%, given the source's predicted $F_{35}$ (Calculated from the best-fit \textsc{dusty} model; see Section 4.5), and the noise level of the observation. Within the rest of our sample of targeted SMC sources, we would expect a number of them to also show OH maser emission. The sample includes oxygen-rich sources slightly less luminous, reddened, and cool, compared to the LMC OH masing sources. We show in Figure \ref{smc_L_vs_J-K} that the SMC sources are typically less reddened at a given luminosity as opposed to the LMC sources, with IRAS 00483$-$7347 again being the exception. Within the figure we have included \textsc{parsec-colibri} stellar isochrones \citep{2017ApJ...835...77M} for different stellar ages that correspond to 2, 4, and 5 solar mass sources for both Magellanic Cloud metallicities\footnote[1]{Ages correspond with the listed mass and metallicity from the \citet{2012MNRAS.427..127B} \textsc{parsec} evolutionary tracks.}. The point within each isochrone where the carbon-to-oxygen ratio becomes greater than 1 is also marked. We see that several of our SMC sources lie upon the 5 M$_{\odot}$ SMC track, where the C/O ratio is still less than 1. These sources have also all been spectroscopically confirmed to be oxygen-rich. We see that the four lower-luminosity sources lie between 2 and 4 M$_{\odot}$ at the SMC metallicity indicating that they are likely carbon stars. We have only plotted isochrones for the lower mass regime as the paths of higher mass isochrones on the HR diagram are more difficult to model and constrain. 

Using the maser efficiency distribution from past maser detections and the noise levels of each targeted SMC observation, we have used a bootstrapping method to predict the likelihood of achieving a maser-pumping efficiency that would result in a $3 \sigma$ level detection (Table \ref{limits}). By replacing each SMC source with a randomly selected maser efficiency from 86 sources in the Galactic and LMC samples from \citet{2017MNRAS.465..403G} with one million iterations, we have calculated the collective probability of not achieving this level in any of our four targeted observation as 0.11\%. This suggests that there is likely an underlying reason for the lack of maser emission in the SMC.

\begin{figure}
\centering
\includegraphics[width=\columnwidth]{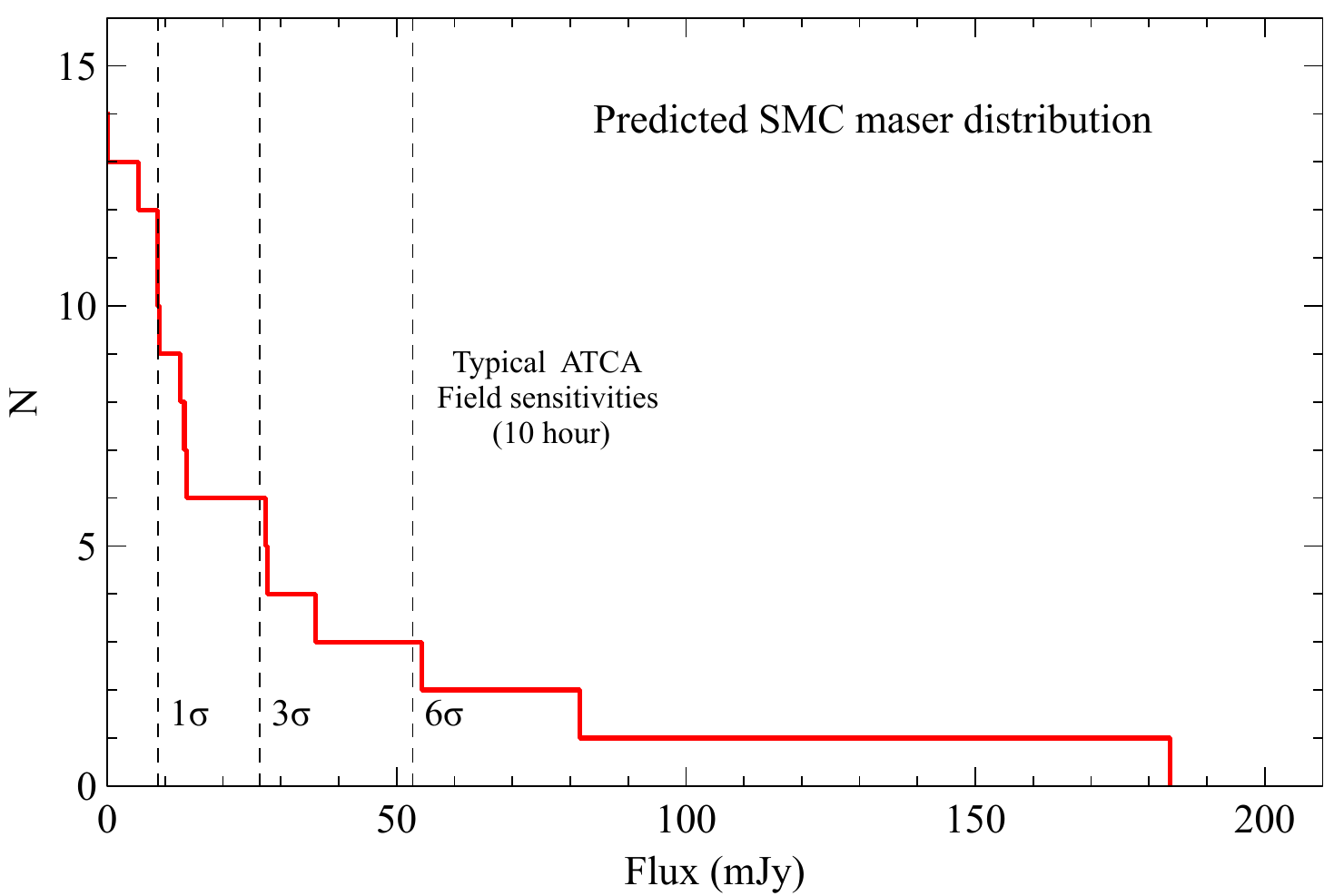}
\caption{The cumulative expected maser peak flux distribution of our SMC sample using the detected LMC maser peak sample from \citet{2017MNRAS.465..403G} and scaling fluxes down by 40\% for distance and another 40\% for metallicity. Also shown are the typical sensitivities at different significance levels of a 10 hour integration with ATCA ($1\sigma\simeq 8.8$ mJy).}
\label{maser_distribution}
\end{figure}

\begin{figure*}
 \centering
 \includegraphics[width=0.49\textwidth]{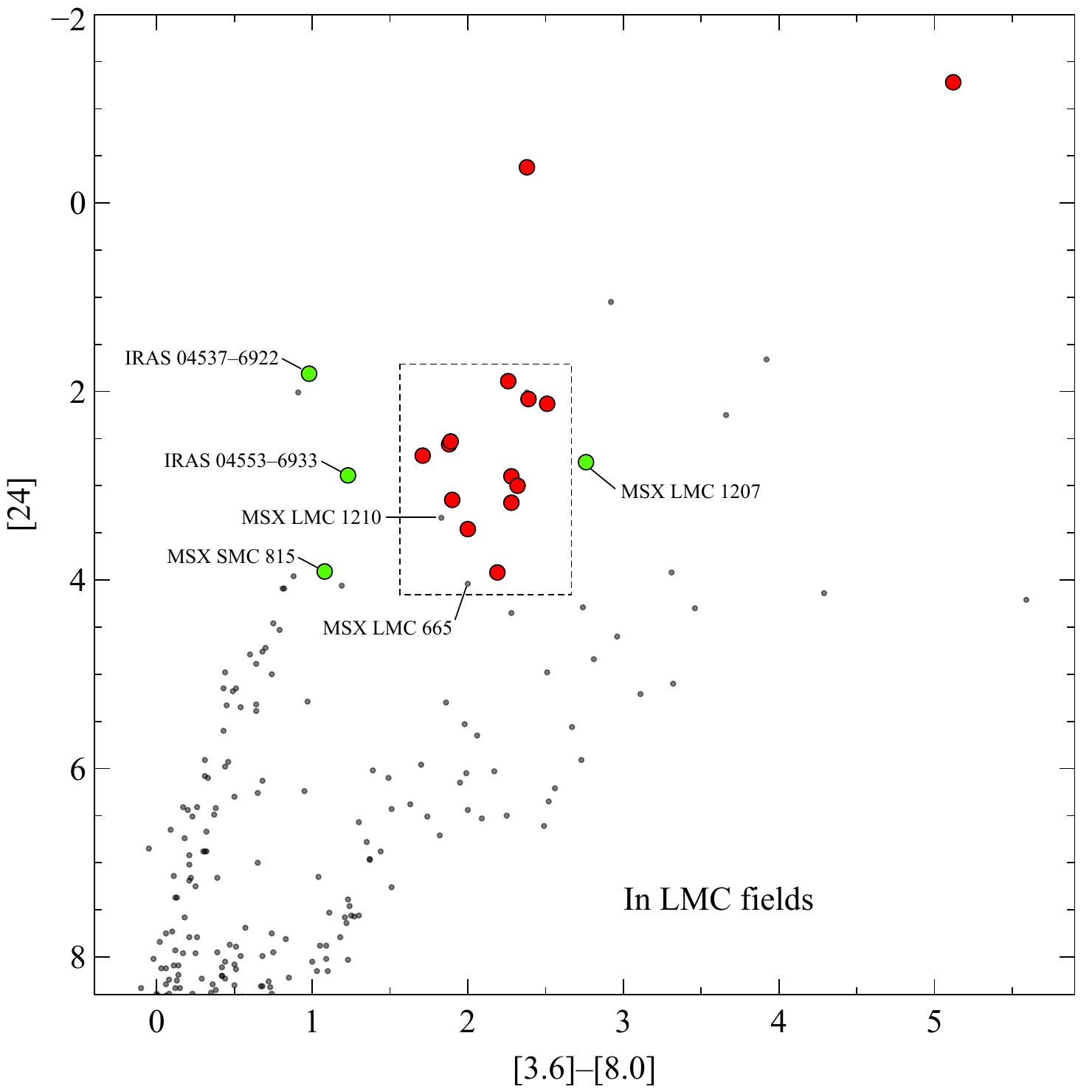}
 \includegraphics[width=0.49\textwidth]{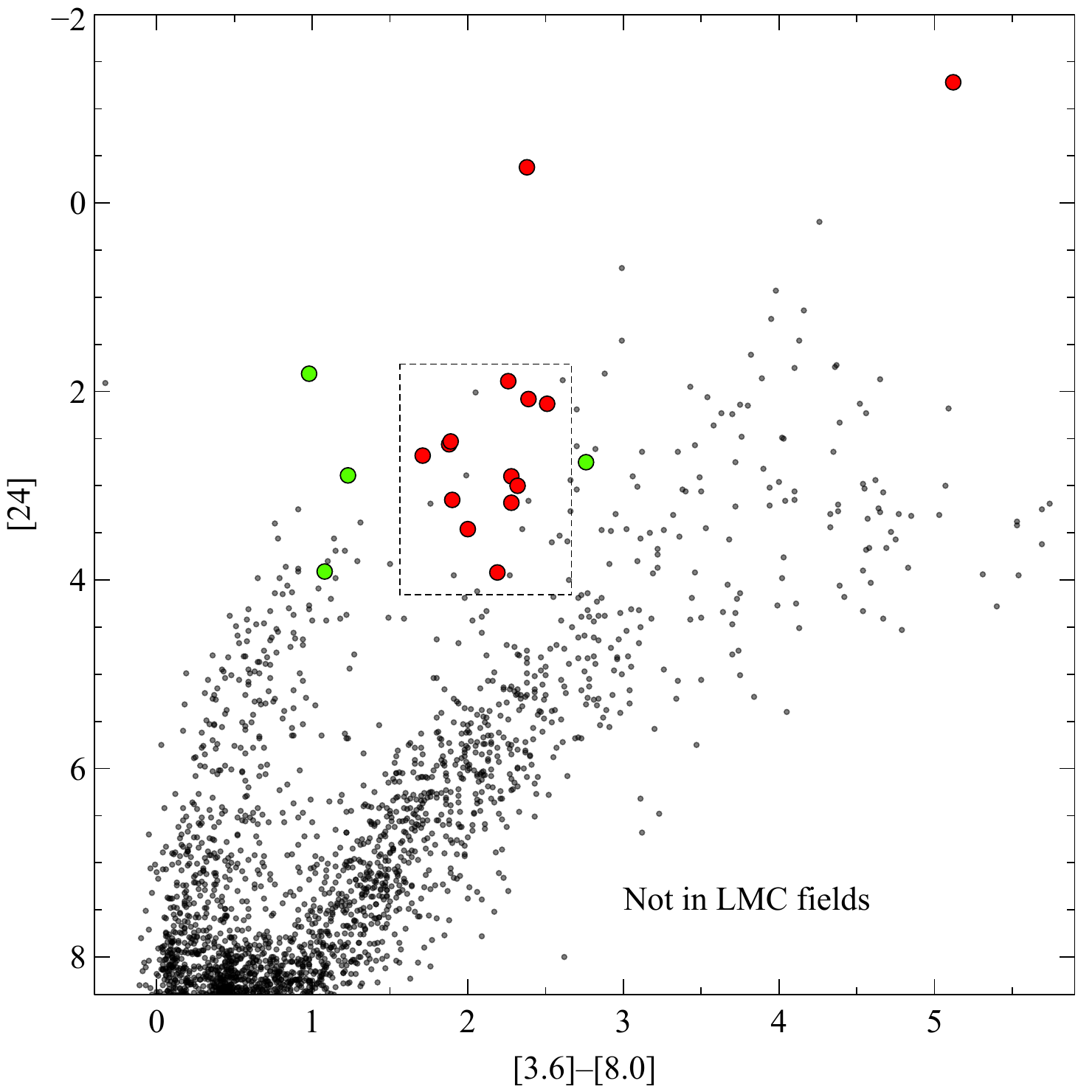}
 \caption[The 24 $\mu$m flux density versus {[3.6]$-$[8.0]} colour for LMC sources in our fields]{The \textit{Spitzer} MIPS 24 $\mu$m flux density versus IRAC [3.6]$-$[8.0] colour for SAGE-LMC sources \textbf{IN} (\textit{Left}) or \textbf{NOT IN} (\textit{Right}) any of our Parkes or ATCA fields targeting OH, with the LMC OH/IR sample in red, LMC non-detections in green, and remaining sources in black. Also shown is the LMC masing zone (dashed box), a region of the CMD typical of OH masing sources.} 
 \label{in_LMC}
\end{figure*}

We have also plotted the maser flux peak distribution expected from the SMC sample (Fig. \ref{maser_distribution}). We have resampled the LMC maser spectra to 0.5 km s$^{-1}$ resolution to match the SMC spectra. The cumulative density of the peak fluxes of the resulting spectra have been used to estimate the SMC peak maser fluxes. We have scaled the fluxes by 40\% for distance, assuming distances of 50 and 60 kpc to the LMC \citep{2013pss5.book..829F} and SMC \citep{2009AJ....138.1661S}, respectively. Assuming the sources are optically thick, the maser emission should depend only on the $F_{35}$ as this value takes metallicity into account. If this is not the case, then the emission will depend on the OH abundance which should scale with metallicity. To compensate for this potential effect we have scaled all LMC peak fluxes down by an additional 40\% for the difference in metallicity. The plot shows that we would have expected 9 sources with peak maser fluxes above the typical noise level of our ATCA observations and 6 sources with peak fluxes above 3$\sigma$. However, this does not account for the difference in star formation history in the Magellanic Clouds. Using values for the star formation rate for the evolution time of $5-6$ M$_{\odot}$ stars (100 Myr) of 3 M$_{\odot}$yr$^{-1}$ and 0.75 M$_{\odot}$yr$^{-1}$ for the LMC and SMC, respectively \citep{2004AJ....127.1531H,2009AJ....138.1243H}, we would expect to have detected one fourth of the sources found in the LMC. This gives us an expected detectable maser population of $1-2$ sources above 3$\sigma$, whereas we detect none.

\begin{figure*}
 \centering
 \includegraphics[width=0.49\textwidth]{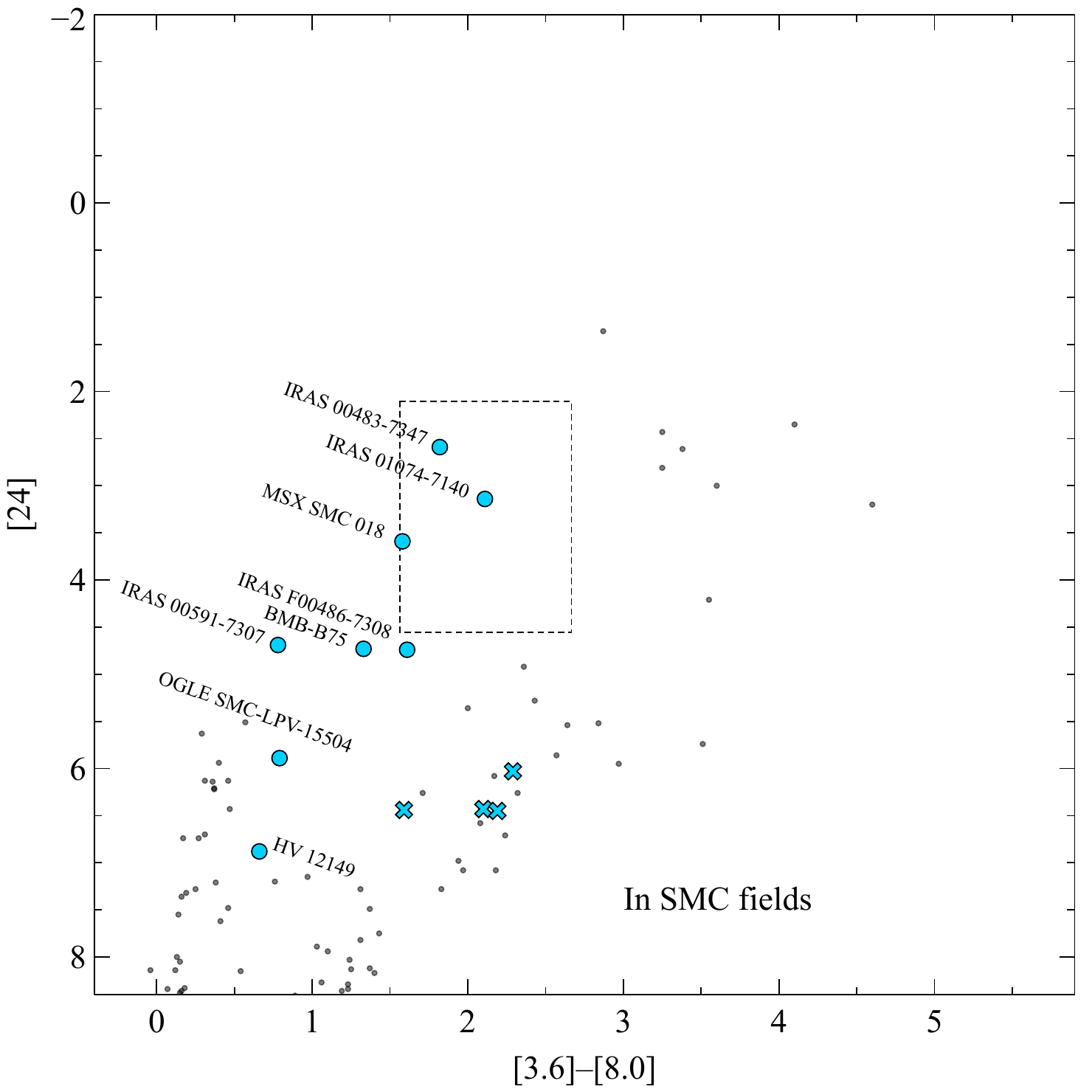}
 \includegraphics[width=0.49\textwidth]{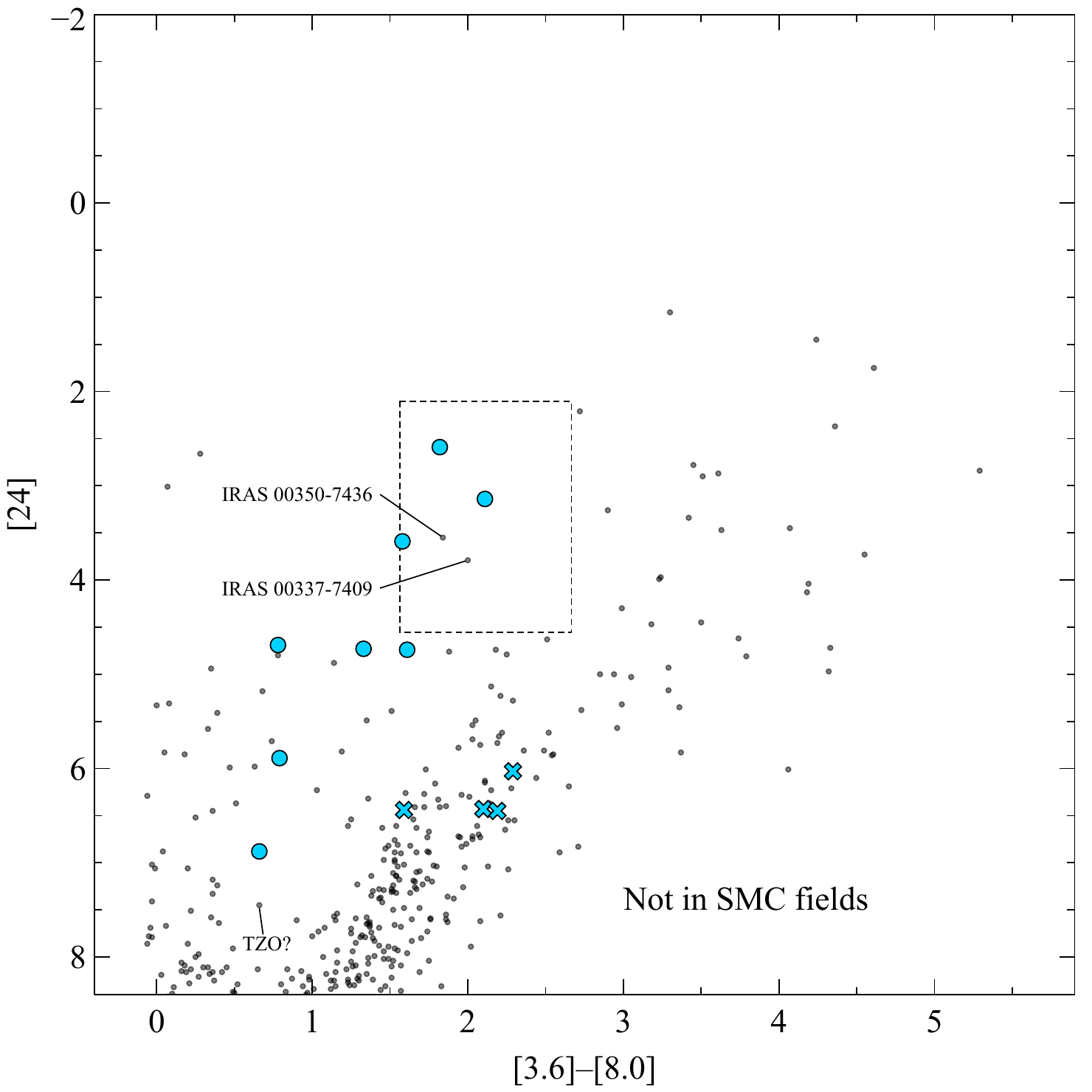}
\caption[The 24 $\mu$m flux density versus {[3.6]$-$[8.0]} colour for SMC sources in our fields]{The \textit{Spitzer} MIPS 24 $\mu$m flux density versus IRAC [3.6]$-$[8.0] colour for SAGE-SMC sources \textbf{IN} (\textit{Left}) or \textbf{NOT IN} (\textit{Right}) any Parkes or ATCA fields targeting OH, with our SMC non-detections as blue circles, our SMC sources likely to be carbon stars as blue crosses, and the remaining sources in black. Also shown is the LMC masing zone (dashed box) scaled down by 40\% (for distance), for the expected SMC OH/IR population. Also shown is the unremarkable position of the Thorne-\.{Z}ytkow or super-AGB candidate-star HV 2112 (TZO?), showing the source is not particularly reddened or luminous at 24 $\mu$m. }
 \label{in_SMC}
\end{figure*}

\subsection{OH/IR populations in the Magellanic Clouds}

\subsubsection{Carbon/Oxygen abundance}
The SMC hosts a smaller population of evolved AGB stars than the LMC, and has a higher fraction of carbon-rich evolved AGB stars \citep{2011AJ....142..103B}. It has also been observed that evolved AGB stars and RSGs in the SMC have less dust than their LMC counterparts \citep{2000A&A...354..125V,2006ASPC..353..211V,vanLoon:2008iv,2010AJ....140..416B}. There are a number of possible explanations for the smaller population of oxygen-rich evolved stars in the SMC, four of which have been proposed by \citet{vanLoon:2008iv}. The first possibility is that oxygen-rich dust grains are more transparent at near-IR wavelengths, which coupled with a smaller dust content in the SMC, have left few obscured oxygen-rich sources. The second possibility is that higher luminosity oxygen-rich sources, which have undergone Hot-Bottom Burning (HBB), are extended, resulting in more diffuse dust. A third possibility is that the size of the SMC, the initial mass function, and the rapid evolution of higher-mass sources, have left few in this evolutionary phase. A final possibility is that as carbon stars form at lower masses within the Magellanic Clouds \citep{1999A&A...344..123M}, this has resulted in fewer lower-mass oxygen-rich sources.

\subsubsection{Warmer evolved red supergiants}
For RSGs, studies have shown that while values for $T_{\textrm{eff}}$ and $M_{\textrm{bol}}$ show agreement between Galactic and LMC sources, sources within the SMC show a much larger spread in $T_{\textrm{eff}}$ for a given $M_{\textrm{bol}}$ \citep{2006ApJ...645.1102L}. This is expected as a result of the enhanced effects of rotational mixing at lower metallicity \citep{2001A&A...373..555M}. In addition to a larger spread in $T_{\textrm{eff}}$, lower metallicity RSGs have been shown to have earlier spectral types \citep{1985ApJS...57...91E,2003AJ....126.2867M}. The expectation is that as spectral types are dependent on Titanium oxide (TiO) bands, and lower metallicity stars will have a lower abundance of TiO, this would result in lower metallicity sources being categorised as earlier spectral types. At the same time, lower metallicity RSGs have also been found to have warmer median $T_{\textrm{eff}}$ \citep{2005ApJ...628..973L,2006ApJ...645.1102L,2009AJ....137.4744L,2007ApJ...660..301M}. As a decrease in the abundance of heavy elements results in a decrease in opacity, and an increase in a star's surface brightness, this results in warmer RSGs at lower metallicity \citep{2013EAS....60..269L}. This would suggest fewer late-type RSGs exist within lower metallicity environments. Yet, observations have shown small populations of late-type RSGs in these lower metallicity environments \citep{1998A&A...332...25G,2007ApJ...660..301M,2007ApJ...667..202L,2010AJ....140..416B}.

\subsection{Isolating the OH/IR sample}

In order to determine the number of sources within the Magellanic Clouds that are capable of OH maser emission, we have considered photometry from the \textit{Spitzer} SAGE-SMC catalogue \citep{2011AJ....142..102G,2011AJ....142..103B}; we have compared these to the SAGE-LMC photometry \citep{2006AJ....132.2034B,2006AJ....132.2268M,2009AJ....138.1003B,2010AJ....140..416B,2009AJ....137.4810S,2010AJ....139.1553V}. 

The sources within the \citet{2017MNRAS.465..403G} LMC OH/IR sample were primarily categorized as Far Infrared Objects (FIR) by \citet{2011AJ....142..103B} except for IRAS 05280$-$6910 which was designated as an extreme or x-AGB star (however \citet{2010A&A...518L.142B} showed that the source is bright at far-IR wavelengths), and IRAS 04553$-$6825, IRAS 05003$-$6712, and IRAS 05558$-$7000, which were not included in the catalogue. In the past, colour-magnitude diagrams (CMDs) have been used to separate different stellar populations within the Magellanic Clouds \citep{2011MNRAS.411.1597W}. Plotting mid-IR colours against mid-IR flux densities, the red supergiants and massive AGB stars create a distinct branch. Looking at the \textit{Spitzer} IRAC [3.6]$-$[8.0] colour against the [24] flux density for the LMC OH/IR sample, the sources occupy a similar high position (or ``masing zone'') on one of these branches (Fig. \ref{in_LMC}). We have plotted two CMDs, one with SAGE-LMC sources that fall within our ATCA and Parkes fields (\textit{left}) and one with sources that do not (\textit{right}). This is done to get an overall idea of the fraction of potential maser sources that have not been covered by our maser survey. We have done this also for our sources in the SMC (Fig. \ref{in_SMC}). 

Within the LMC sample, there are two sources in our fields that have been identified as AGB stars and lie within the LMC masing zone, but do not exhibit circumstellar maser emission. Within our field are the sources MSX LMC 1210 and MSX LMC 1207, sources that do not exhibit maser emission. The first source, MSX LMC 1210, with a $J-K$ colour of 2.94 mag, is bright at mid-IR ($F_{24}=387$ mJy), and has a pulsation period of 1050 d \citep{2009AcA....59..239S}. The second source, IRAS 05287$-$6910, is also  bright in the mid-IR ($F_{24}$=1.1 Jy), exhibits silicate in emission \citep{2011ApJS..196....8L}, and lies in the cluster NGC 1994 \citep{2005A&A...442..597V}. The cluster has an age of 11.5 Myr \citep{1988AJ.....96.1383E} and a metallicity of 0.58 Z$_{\odot}$\citep{1998A&A...332...46O}.

Within the masing zone of sources not within the \citet{2017MNRAS.465..403G} LMC fields are three blue supergiants, three young stellar objects, four emission line stars, two carbon stars, a planetary nebula, and two sources that would be good targets for OH maser emission. The first source, IRAS 04523$-$7043, has a 24 $\mu$m flux density of 3.59 mag, and a pulsation period of 890 d \citep{2009AcA....59..239S}. The second source, 2MASS J05241334$-$6829587, has a 24 $\mu$m flux density of 2.58, a $J-K$ colour of 3.2 mag, a pulsation period of 899 days \citep{2011yCat..35360060S}.

Within our SMC fields, there are three target sources that lie within the scaled LMC masing zone in the CMD: IRAS 00483$-$7347, IRAS 01074$-$7140, and MSX SMC 018 (Fig. \ref{in_SMC}). To the right of the region are several young stellar objects and emission line stars \citep{2013ApJ...778...15S}. There are also two sources that do not lie in any of our observed fields, IRAS 00350$-$7436 and IRAS 00337$-$7409 (HD 3407) within this same masing region. IRAS 00350$-$7436 has been found to be the most luminous carbon-rich object in the SMC \citep{1989MNRAS.238..769W,Srinivasan:2016jp}. In the past the source has been categorized as a carbon star \citep{2015MNRAS.451.3504R}, a post-AGB star \citep{2005A&A...434..691M} and an interacting binary star \citep{1989MNRAS.238..769W}. The second source, IRAS 00337$-$7409, is a foreground star with a radial velocity of 58 km s$^{-1}$ \citep{1972POStr...2....1F}. No other sources within the SAGE-SMC survey lie within the masing zone as defined by the LMC OH/IR sources. This gives us confidence that we have surveyed all SMC sources that we would expect to exhibit circumstellar OH maser emission at our current detection thresholds. Also shown in the figure is the Thorne-\.{Z}ytkow Object candidate star HV 2112 \citep{2014MNRAS.443L..94L,2014MNRAS.445L..36T}. These stars, while yet to be confirmed observationally, are thought to be red giants or supergiants with neutron stars at their cores. Alternatively, it has been suggested it could be a super-AGB star \citep{1983ApJ...272...99W,1990ApJ...361L..69S,2014MNRAS.445L..36T}, or even a Galactic foreground S-type star \citep{2016MNRAS.458L...1M}, though its SMC membership seems secure on the basis of the radial velocity of $v_{\rm rad}=150$ km s$^{-1}$ \citep{2015A&A...578A...3G} and proper motion \citep{2016MNRAS.459L..31W}. The position of this source, with respect to the other SMC sources, suggests it is not particularly reddened or luminous at 24 $\mu$m.

\subsection{Maser strength}

Now that we have established the expectation of maser detections, we will aim to understand their absence. Several conditions must be met in order to produce circumstellar OH maser emission. First, as the transition is a population inversion, a dense population of OH with number densities typically $10^{12}-10^{16}$m$^{-3}$ is required \citep{2006evn..confE..42R}. The masing environment must have a sufficiently long path length to provide the high column density, needed to amplify the maser \citep{2005MNRAS.364..783G}. Masers can also only occur in a certain radius range of the parent star. If masers occur inside a radius of around several hundred au from the star, densities are so high that collisional pumping rapidly increases and ``quenches'' the maser. Outside a few thousand au, lower densities are not enough to sustain the pump rate \citep{1992ASSL..170.....E,Vlemmings2006}. In addition to having a suitable masing environment, significant IR flux must be present to pump the maser. It is possible that circumstellar environments in the SMC do not adhere to these conditions.

\subsubsection{OH abundance}

Within oxygen-rich evolved stars, the abundant oxygen allows for the formation of a variety of minerals, CO, and a significant amount of water. A smaller fraction of water has been found around carbon stars \citep{2001Natur.412..160M,2010Natur.467...64D} likely a result of sublimating comets or the Fischer-Tropsch catalysis mechanism which converts CO and H$_{2}$ into hydrocarbons and water using iron as a catalyst \citep{2004ApJ...600L..87W}. In oxygen-rich stars water molecules are transported by stellar pulsation shocks, and the dust-driven wind, out to typical distances of $20-80$ au \citep[e.g.][and references therein]{2002PASJ...54..757S}. This water is then photodissociated by interstellar UV radiation, creating OH. The abundance of OH will depend on the initial abundance of water but also the efficiency of converting water into OH, i.e. the penetrating strength of the UV radiation within the circumstellar envelope \citep{1994A&A...292..371L}.

The observations and modeling of the conditions in metal-poor dwarf galaxies and nearby globular clusters indicate a much stronger radiation field and more porous molecular gas in lower metallicity environments \citep{2006A&A...446..877M,2015MNRAS.453.4324M,2015MNRAS.448..502M,2017MNRAS.464.1512W}. With more ionized hydrogen and less interstellar dust, the expectation is that the lower opacity will have an impact on chemical abundances in circumstellar environments. This may also be the case with our SMC sources, yet several LMC sources that also lie within clusters (e.g. IRAS 05280$-$6910 and IRAS 05298$-$6957) seem unaffected by their cluster environment \citep{2002math.ph...1001Q,2005A&A...442..597V,1992ApJ...397..552W}.

The source of OH in evolved stars is water, and an understanding of the abundance of water can inform our understanding of the abundance of OH. Past surveys have targeted 22-GHz water maser emission in the SMC \citep{2013MNRAS.432.1382B}, and while interstellar maser emission has been detected in star forming regions, no circumstellar maser emission has been confirmed. All of our brightest SMC sources (IRAS 00483$-$7347, IRAS F00486$-$7308, IRAS 00591$-$7307, IRAS 01074$-$7140, BMB-B75, and MSX SMC 018) have shown water in absorption at $\sim$ 3 $\mu$m \citep{vanLoon:2008iv}. It was suggested that this may be indicative of an advanced evolutionary stage, yet they also note that the comparison LMC sample does not show strong absorption from these molecular bands. They go on to suggest that it may be possible that less of the water molecules are being locked up in dust or that the population of water molecules may be closer to the stellar surface, increasing the column density as well as the excitation temperature. This result may hint at a potential difference in the circumstellar environments of our LMC and SMC OH/IR samples. However, the lack of OH maser emission in these sources is not due to a lack of water within their circumstellar environments. A deep survey for water maser emission in our top maser candidates may provide a better opportunity to observe circumstellar maser emission in the SMC.

It is highly possible that a denser circumstellar wind will result in a higher water column density. This higher density would create a more extended water masing region and result in a higher water maser flux density. With water closer to the star, it is less susceptible to external forces like interstellar UV radiation that convert water to OH. Also unlike 1612-MHz OH maser emission, water masers are pumped by both radiation and collisions with H$_2$ molecules \citep{2006A&A...447..949B}. It is possible that the relative strengths of the maser pumping mechanisms may differ at lower metallicity. With less dust, and as a result lower mid-IR flux, collisional pumping may be more effective at lower metallicity. These facts have led us to the conclusion that the chances of detection in the SMC may favour water over OH masers.

\subsubsection{Maser pumping}

\begin{figure}

 \includegraphics[width=\columnwidth]{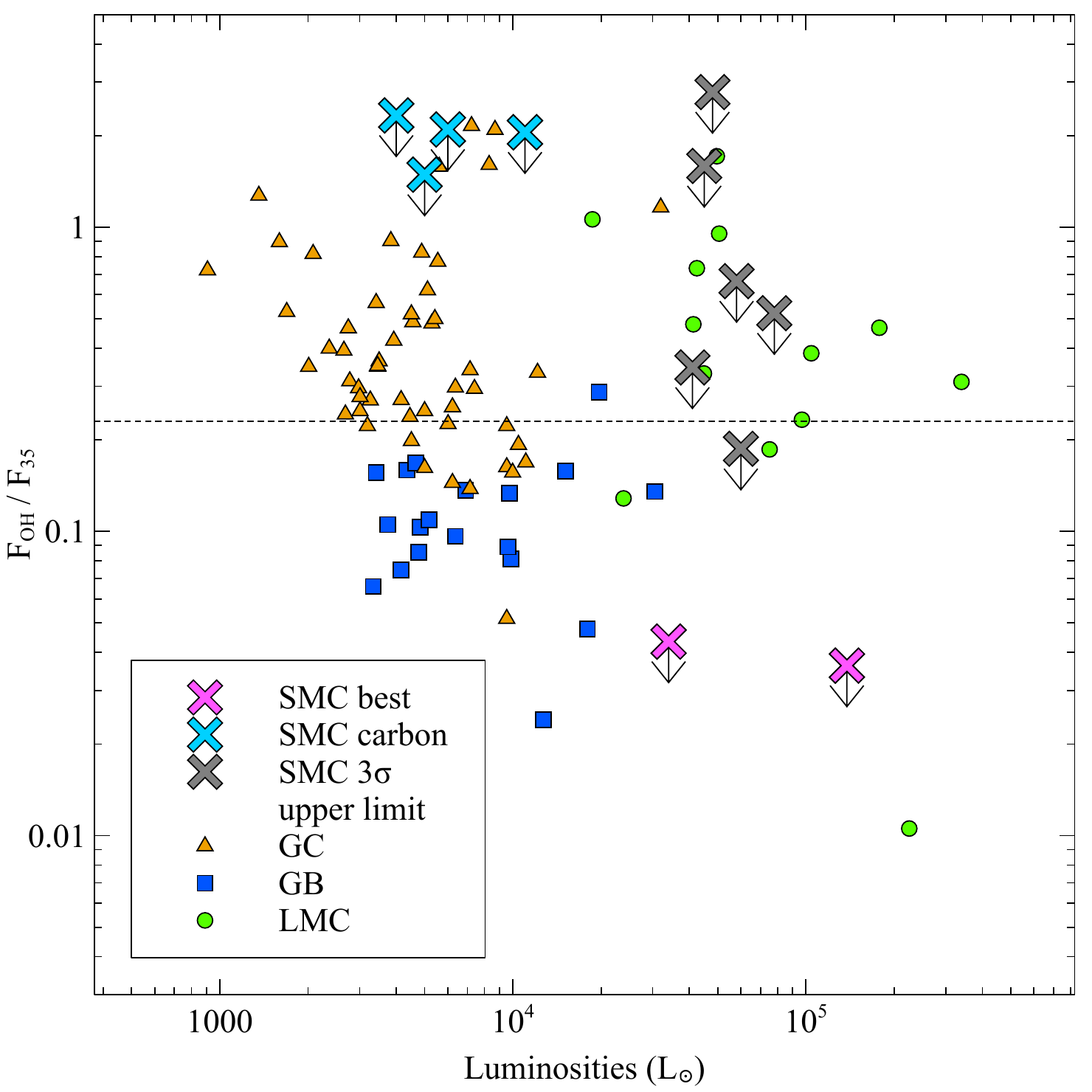}
 \caption[Expected maser efficiency]{Maser efficiency of converting 35 $\mu$m photons to 1612-MHz maser photons for the Galactic Centre (GC), Galactic bulge (GB), and LMC samples from \citet{2017MNRAS.465..403G}. The dotted line represents the median efficiency for the three samples of $23\%$. Also plotted are the 3$\sigma$ upper limits for the maser efficiencies of the SMC sample, our best two candidates: IRAS 01074$-$7104 and IRAS 00483$-$7308 (SMC best), and the SMC sources likely to be carbon stars (SMC carbon), taken from the ratio of the predicted 35 $\mu$m flux to three times the observation noise level.}
 \label{maser_efficiency}
 \end{figure}

The 1612-MHz maser is a product of a population inversion that results from the pumping by infrared photons of a transition at 35 $\mu$m. As a result, the strength of maser emission will depend on the dominant source of IR emission, circumstellar dust. The theoretical maximum pumping efficiency for the 1612-MHz maser transition is 25\% \citep{1992ASSL..170.....E}. This assumes the maser is saturated and the maser emission is isotropic. However, results from \citet{2005A&A...434..201H} have shown stars with pump rates as low as 0.05, 1/5 of this value. We see a large spread in pumping efficiency within the LMC and Galactic samples that extends far above this theoretical maximum with a median maser efficiency of 23\% (Fig. \ref{maser_efficiency}). We expect that as the radio and IR measurements were taken at different times, and these sources are highly variable, this has contributed to the large scatter. We have calculated median maser efficiencies for the Galactic Centre, Galactic bulge and LMC OH/IR samples from \citet{2017MNRAS.465..403G} of 34.8\%, 10.5\%, 38.5\%, respectively (see below). It should be noted that these are derived from spectra spanning a range of velocity resolutions between $\sim 0.5-1.5$ km s$^{-1}$, which will introduce some uncertainty. The much lower maser efficiency in the Galactic Bulge sample is surprising as SED fitting has found typically lower gas-to-dust ratio, and dramatically higher optical depth in dust emission than in the Galactic Centre and LMC samples \citep{2017MNRAS.465..403G}. This implies a larger supply of IR photons per OH molecule in the Galactic Bulge sources, yet we see a lower maser efficiency.

Figure \ref{maser_efficiency} shows the relationship of maser efficiencies and luminosities of the Galactic Centre, Galactic Bulge and LMC OH/IR samples from \citet{2017MNRAS.465..403G}. We have used predicted 35 $\mu$m flux densities from the best-fit \textsc{dusty} model of the sources (further explained in Section 4.5). The OH maser flux densities ($F$\textsubscript{OH}) are the peak flux density of both maser peaks combined for the Galactic samples from \citet{2015A&A...582A..68E}, and LMC sample from \citet{2017MNRAS.465..403G}. Also shown in this figure, are the upper limits of the maser efficiency of the SMC non-detections, as well as the two sources that we consider our best maser candidates IRAS 00483$-$7347 and IRAS 01074$-$7104, calculated using the ratio of the predicted 35 $\mu$m flux densities from the best-fit \textsc{dusty} model of the sources (further explained in Section 4.5) and three times the noise level. Our two best maser candidates are clearly below the expected maser efficiencies of all other sources of comparable luminosities except for the unique LMC RSG IRAS 05280$-$6910, which is suspected to exhibit a unique geometry \citep{2017MNRAS.465..403G}. This again suggests that, while not the case in the LMC, the lower metallicity of the SMC may be a critical factor in determining the presence of maser emission. These pumping efficiencies are comparable to those found in \citet{2005A&A...434..201H}, yet we have expressed our values in terms of the peak intensity and not photon flux density, and our comparative analysis between the galaxy and the Magellanic Clouds still points to the expectation of detection within our SMC sample. 

\begin{table*}
\centering
\caption[Results of spectral energy distribution fitting for the SMC sample]{The results of SED fitting for the SMC sample. The values are the modelled results from the best-fitting \textsc{dusty} models with luminosities ($L$), effective and inner dust temperatures ($T$\textsubscript{eff,DUSTY}, $T$\textsubscript{inner}), optical depths specified at 10 $\mu$m ($\tau$), and gas mass-loss rates ($\dot{M}$); mass-loss rates are calculated with the assumption of an SMC gas-to-dust ratio of 1000 \citep{1998AJ....115..605L}. A discussion of the expected errors can be found in the text. Also listed are the predicted expansion velocities (v\textsubscript{exp,Goldman}) using our new mass-loss prescription from \citet{2017MNRAS.465..403G} assuming a metallicity of one fifth solar. } 
\begin{minipage}{\textwidth}
\centering
\begin{tabular}{ l c c c c c c c}
  \hline 
Object & 
Grain &
$L$ & 
v\textsubscript{exp,Goldman} & 
$T$\textsubscript{eff,DUSTY} &
$T$\textsubscript{inner} &
$\tau$   &
$\dot{M}$
\\
name & 
type &
($10^3$ L\textsubscript{$\odot$}) &
(km s$^{-1})$ & 
(K)&
(K)&
&
(M\textsubscript{$\odot$} yr$^{-1}$) 
\\
\hline  
\textit{Parkes targets}\\ 
IRAS\,00483$-$7347 & 
O &
\llap{13}8 & 
5.4 &
3100 &
\llap{1}200 &
1.3\rlap{1} &
1.8 $\times 10^{-4}$ \\
IRAS\,F00486$-$7308 &
O &
\llap{4}1 & 
3.3 &
3700 &
\llap{1}000 &
0.2\rlap{7} &
3.9 $\times 10^{-5}$ \\
IRAS\,00591$-$7307 &
O &
\llap{7}8 & 
4.3 &
2800 &
600 &
0.1 &
3.4 $\times 10^{-5}$ \\
IRAS\,01074$-$7140&
O &
\llap{3}4 &
3.1 &
3100 &
350 &
0.4\rlap{8} &
5.9 $\times 10^{-5}$ \\
\\
\textit{ATCA sources} \\ 
OGLE SMC-LPV-15504 &
O &
\llap{4}5 &
3.4 &
2700 & 
\llap{1}000 &
0.1\rlap{2} &
1.5 $\times 10^{-5}$
\\
2MASS J01033142$-$7214072 &
C &
4 &
1.3 &
3700 &
\llap{1}000 &
0.2\rlap{0} &
1.2 $\times 10^{-5}$
\\
&
O &
7 &
1.6 &
3000 &
\llap{1}400 &
1.4\rlap{9} &
1.8 $\times 10^{-5}$
\\
HV 12149 &
O &
\llap{4}8 &
3.5 &
3000 &
\llap{1}400 &
0.1 &
1.0 $\times 10^{-5}$ 
\\
OGLE SMC-LPV-14322 & 
C &
\llap{1}1 &
2.0 &
3400 &
\llap{1}400 &
0.2\rlap{0} &
2.0 $\times 10^{-5}$ 
\\
&
O &
\llap{1}1 &
2.0 &
3000 &
\llap{1}400 &
1.0\rlap{9} &
2.0 $\times 10^{-5}$ 
\\
2MASS J00592646$-$7223417 &
C &
5 &
1.5 &
3500 &
\llap{1}000 &
0.2\rlap{4} &
1.8 $\times 10^{-5}$ 
\\
&
O &
8 &
1.7 &
3000 &
\llap{1}400 &
1.3\rlap{1} &
1.8 $\times 10^{-5}$ 
\\
BMB-B75 &
O &
\llap{5}8 & 
3.8 &
3400 &
600 &
0.1 &
3.5 $\times 10^{-5}$ 
\\
OGLE J004942.72$-$730220.4 &
C &
6 &
1.5 &
2800 &
800 &
0.1\rlap{5} &
1.6 $\times 10^{-5}$ 
\\
&
O &
8 &
1.7 &
3000 &
\llap{1}400 &
1.8 &
2.2 $\times 10^{-5}$ 
\\
MSX SMC 018 &
O &
\llap{6}0 &
3.8 &
2700 &
600 &
0.3\rlap{5} & 
7.0 $\times 10^{-5}$
\\ \hline
\label{table_smc_sed_results}
\end{tabular}
\end{minipage}
\end{table*}

\begin{figure*} 
 \centering
 \captionsetup[subfigure]{labelformat=empty}
 \subfloat[]{\includegraphics[width=0.49\textwidth,height=0.49\textheight,keepaspectratio]{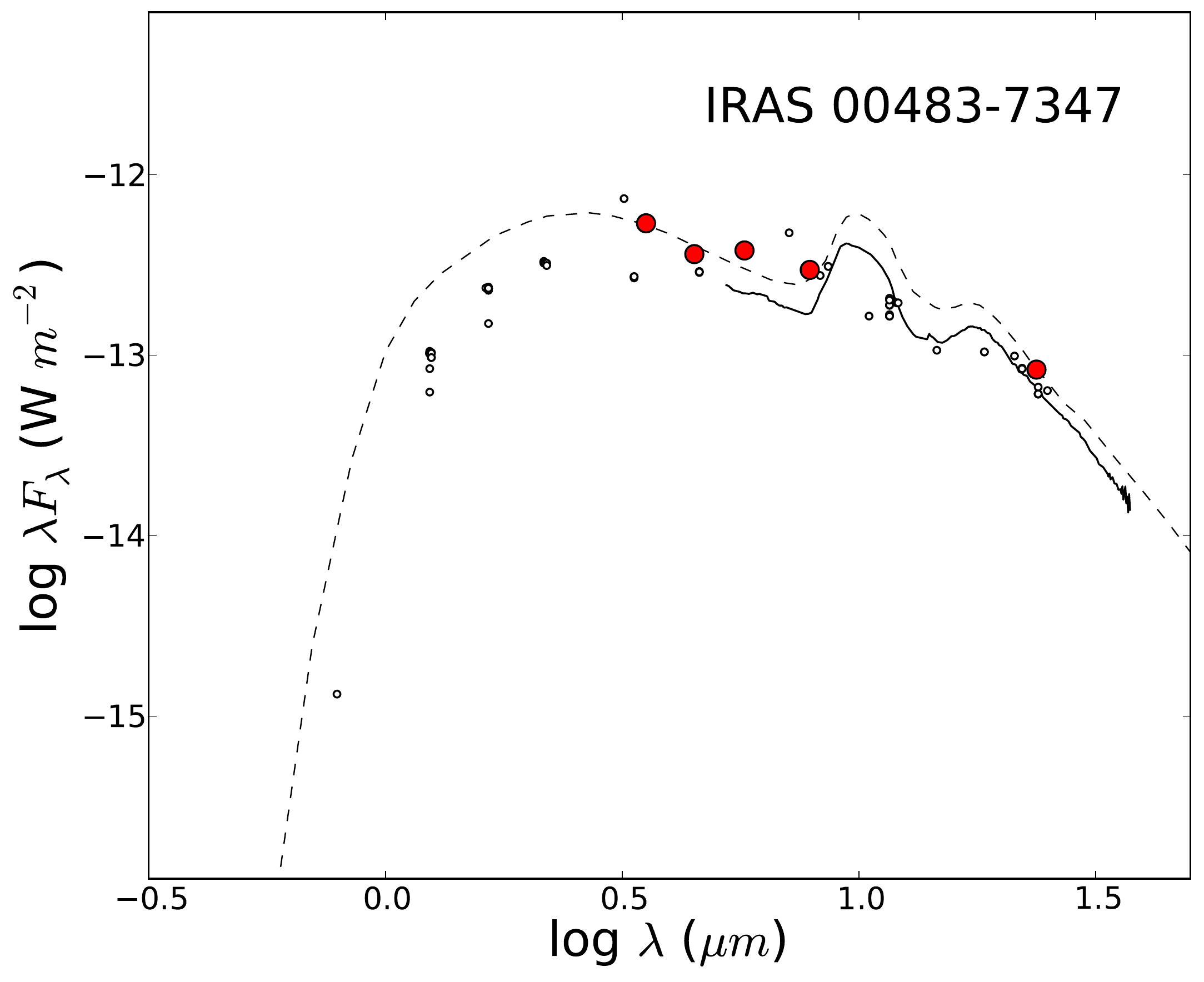}}
 \subfloat[]{\includegraphics[width=0.49\textwidth,height=0.49\textheight,keepaspectratio]{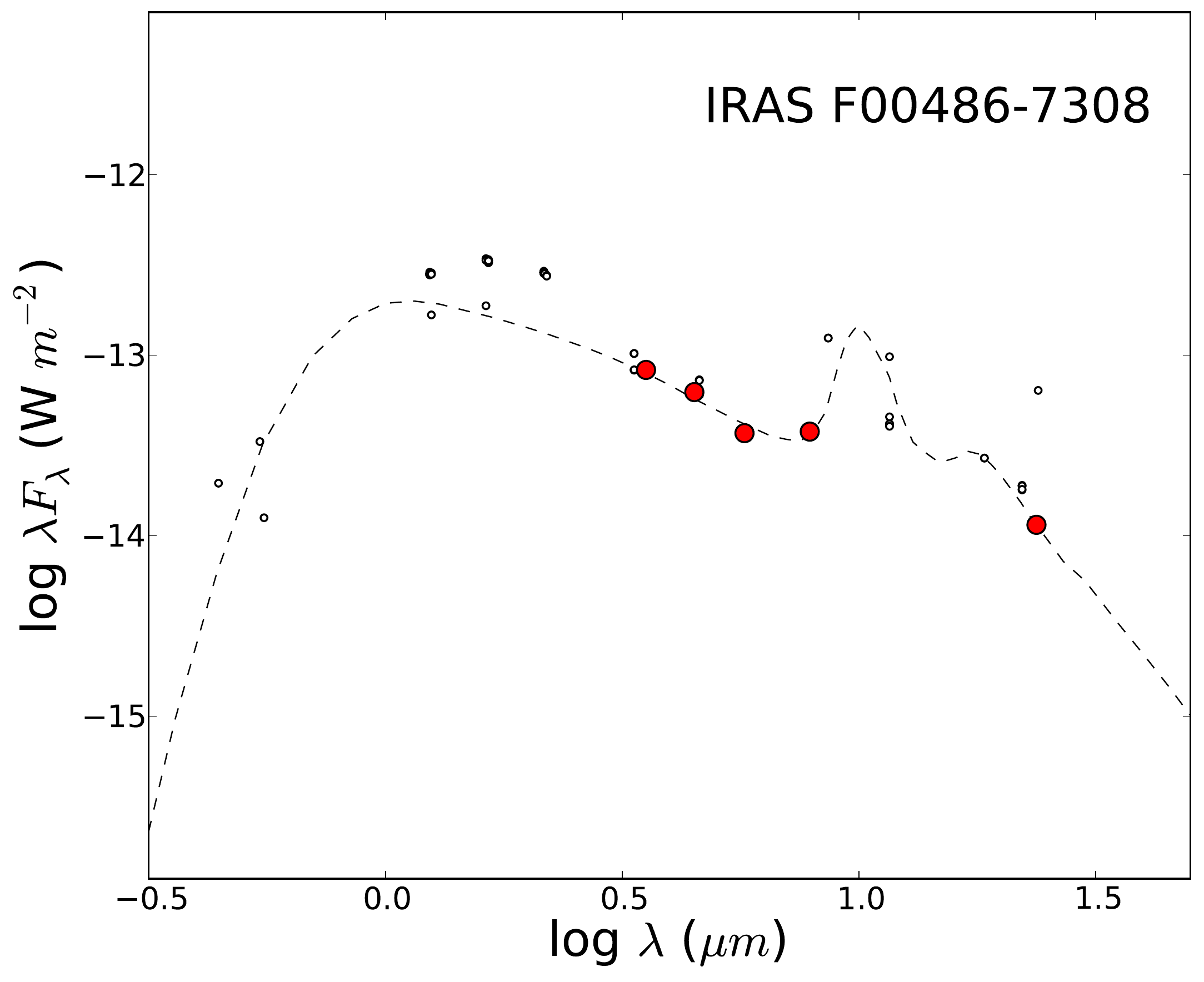}}\\ \vspace{-0.84cm}
 \subfloat[]{\includegraphics[width=0.49\textwidth,height=0.49\textheight,keepaspectratio]{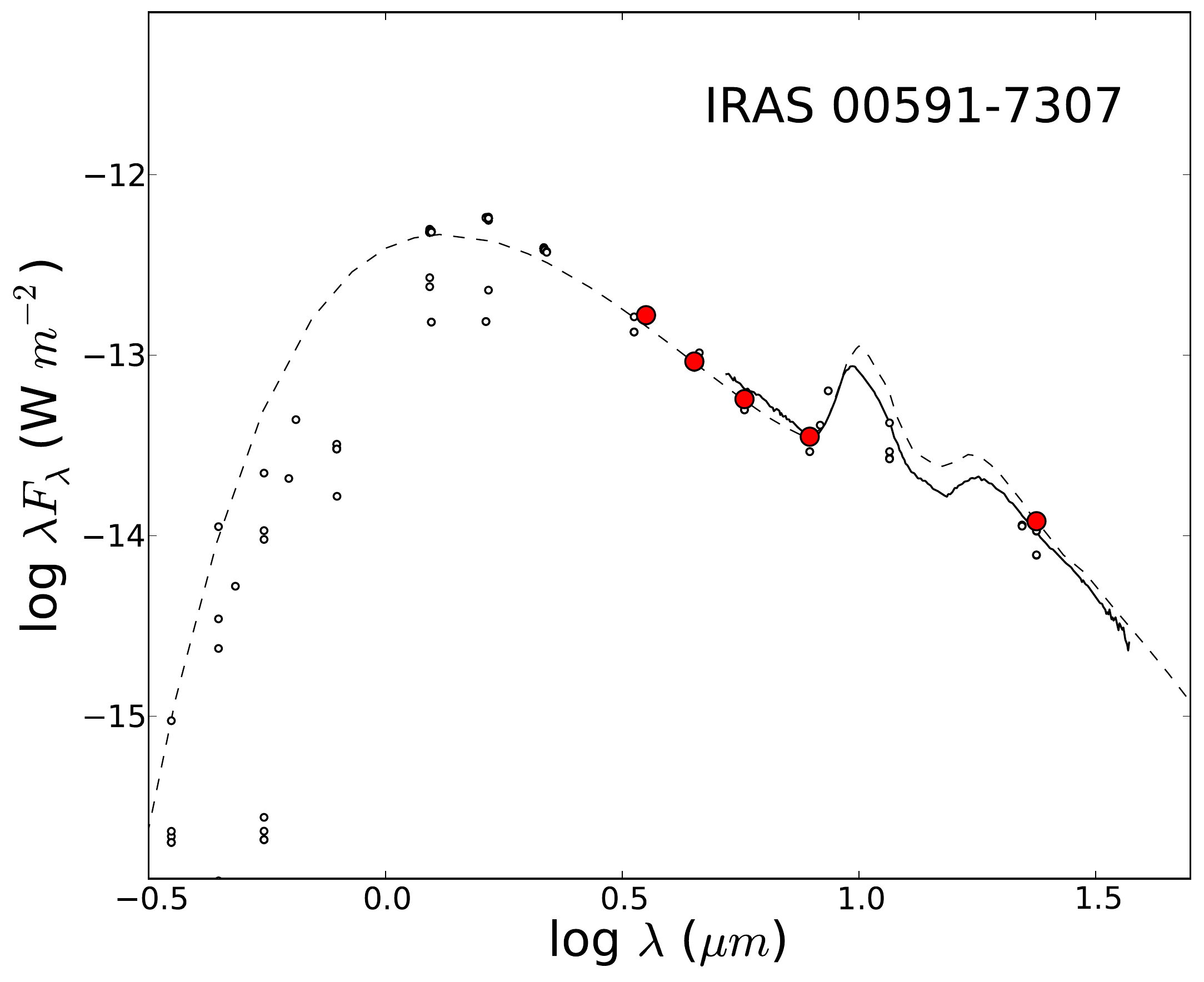}}
 \subfloat[]{\includegraphics[width=0.49\textwidth,height=0.49\textheight,keepaspectratio]{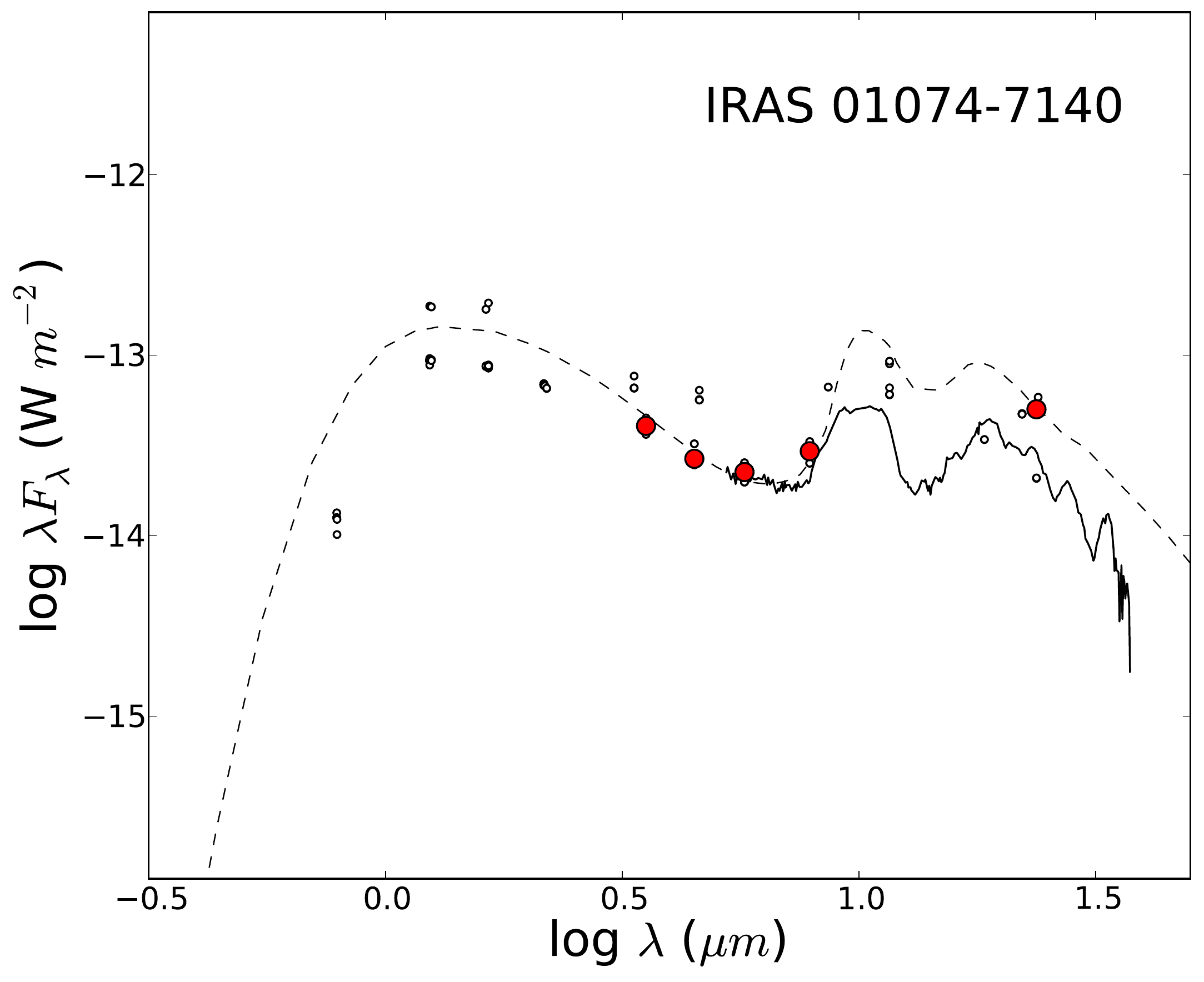}}\\ \vspace{-0.84cm} 
 \subfloat[]{\includegraphics[width=0.49\textwidth,height=0.49\textheight,keepaspectratio]{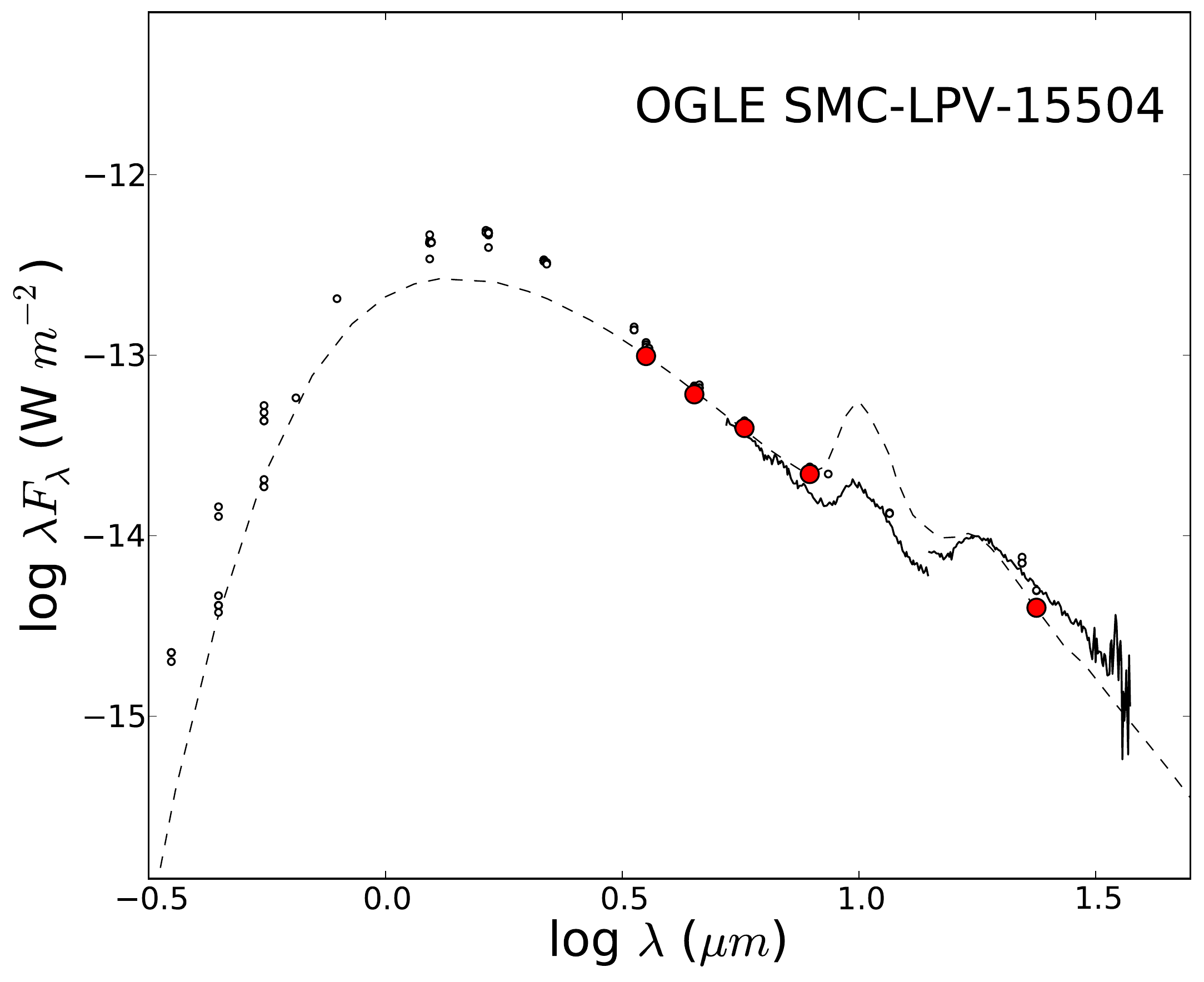}}
 \subfloat[]{\includegraphics[width=0.49\textwidth,height=0.49\textheight,keepaspectratio]{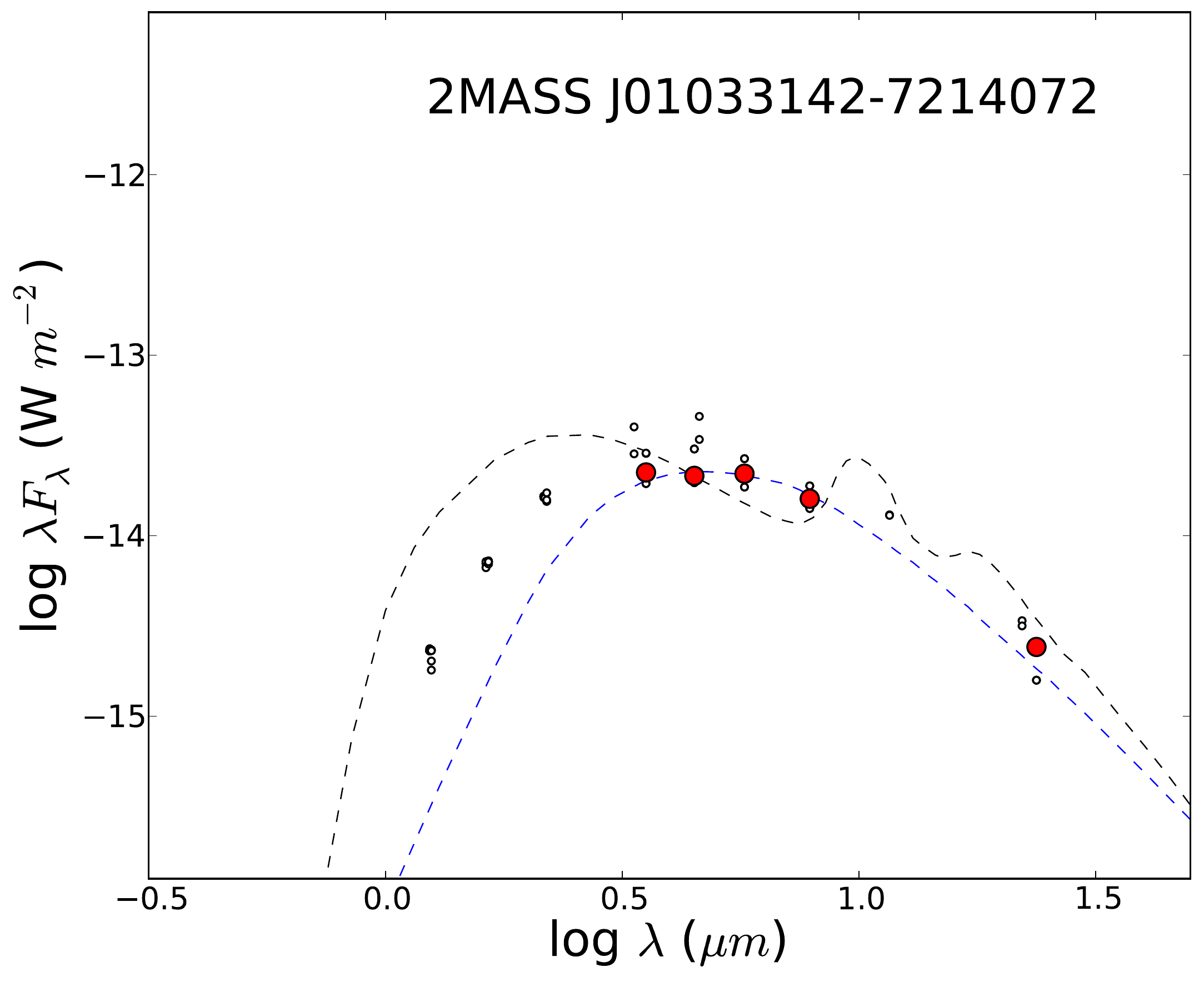}}\\ 
 \caption[The SMC spectral energy distributions]{The SED fitting of \textsc{dusty} models to \textit{Spitzer} IRAC and MIPS photometry of SMC sources (shown in red), with our best-fitting model with oxygen-rich grains (dashed line), \textit{Spitzer} IRS spectra (in solid black), and the remaining available photometry (shown with small open circles). Also shown is our best-fitting model fit with carbon-rich dust grains for our sources that are likely carbon stars (in blue).}
 \label{smc_seds}
 \end{figure*}

\begin{figure*}
 \centering
 \captionsetup[subfigure]{labelformat=empty}
 \ContinuedFloat
 \subfloat[]{\includegraphics[width=0.49\textwidth,height=0.49\textheight,keepaspectratio]{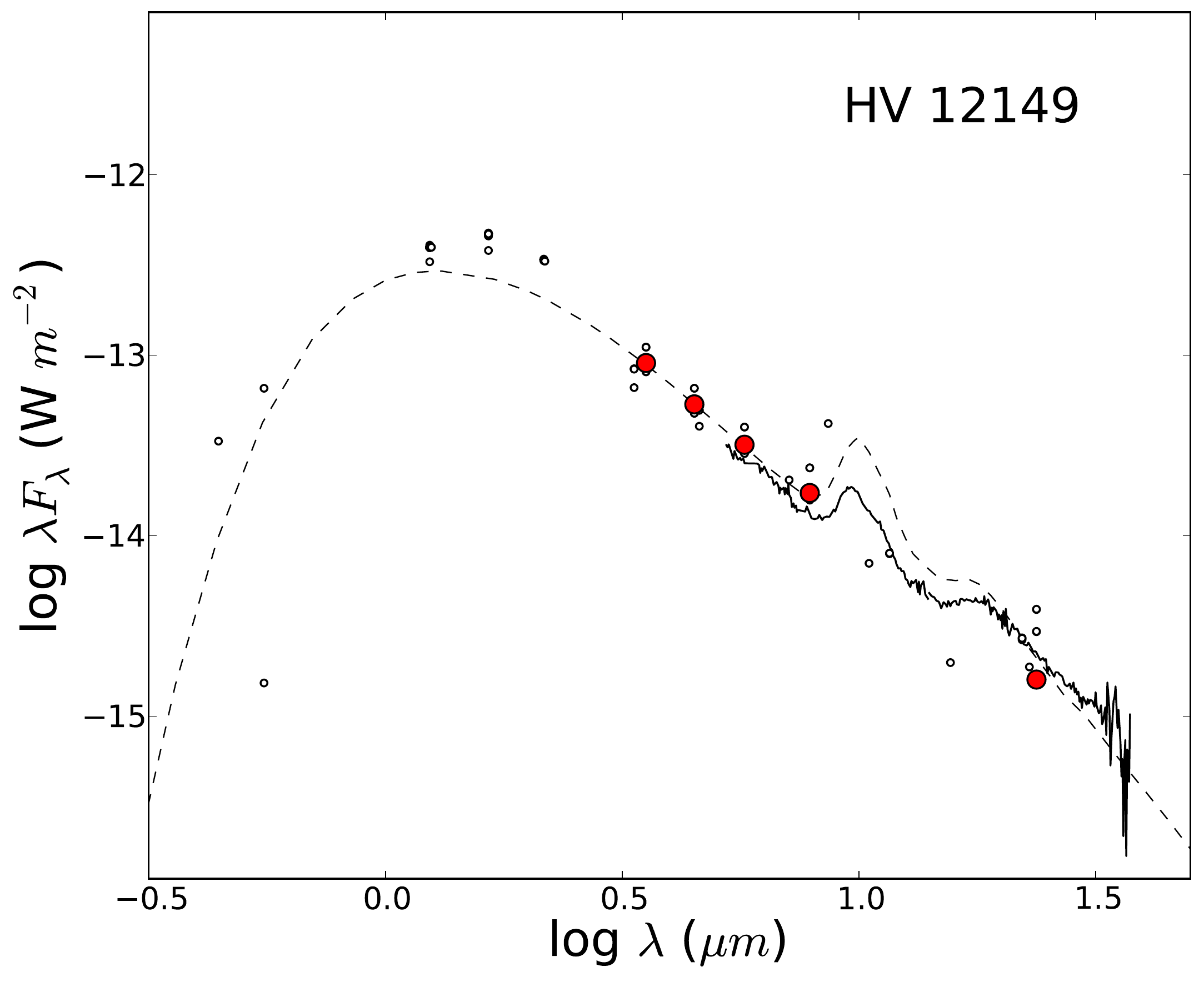}}
 \subfloat[]{\includegraphics[width=0.49\textwidth,height=0.49\textheight,keepaspectratio]{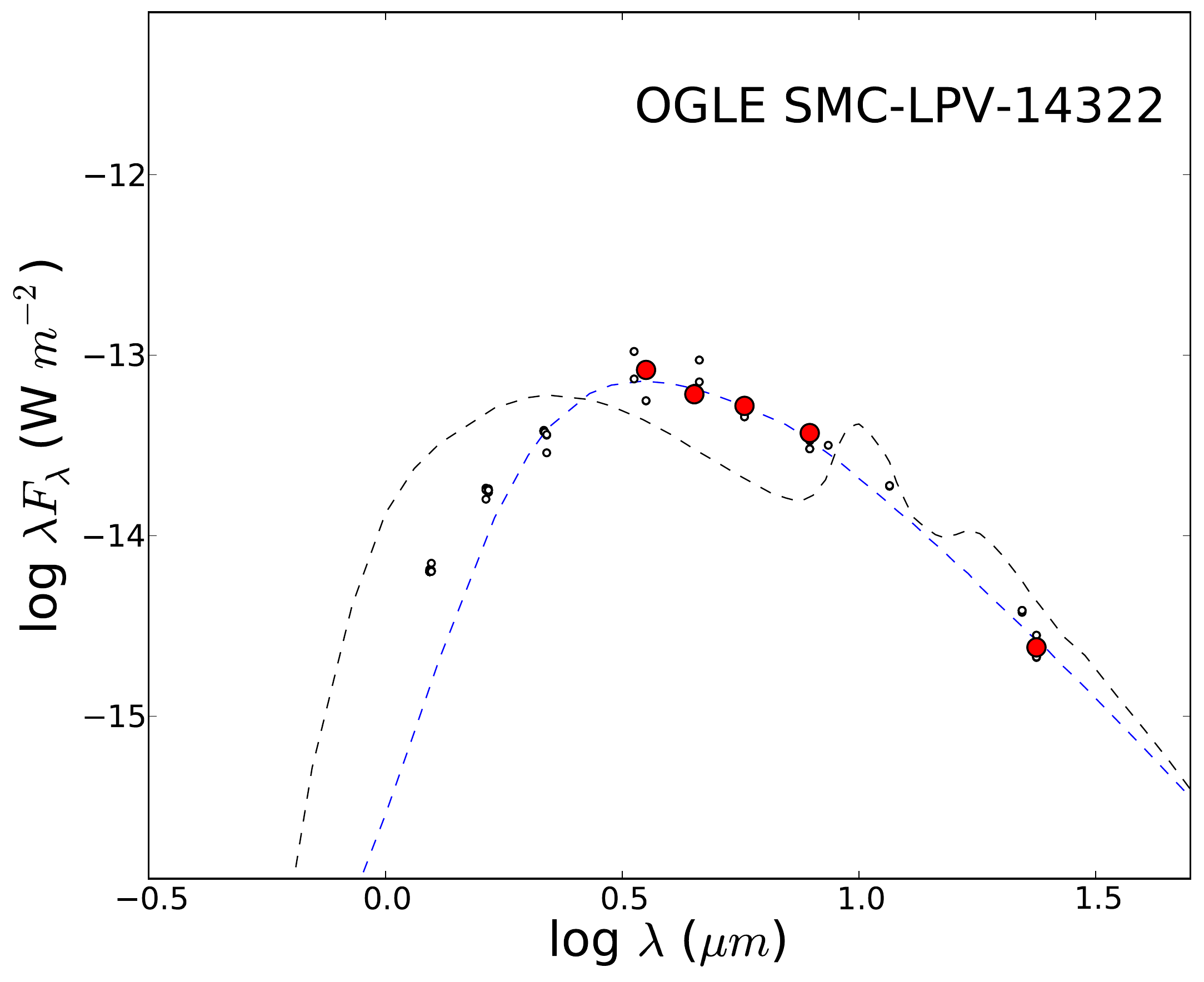}}\\ \vspace{-0.5cm}
 \subfloat[]{\includegraphics[width=0.49\textwidth,height=0.49\textheight,keepaspectratio]{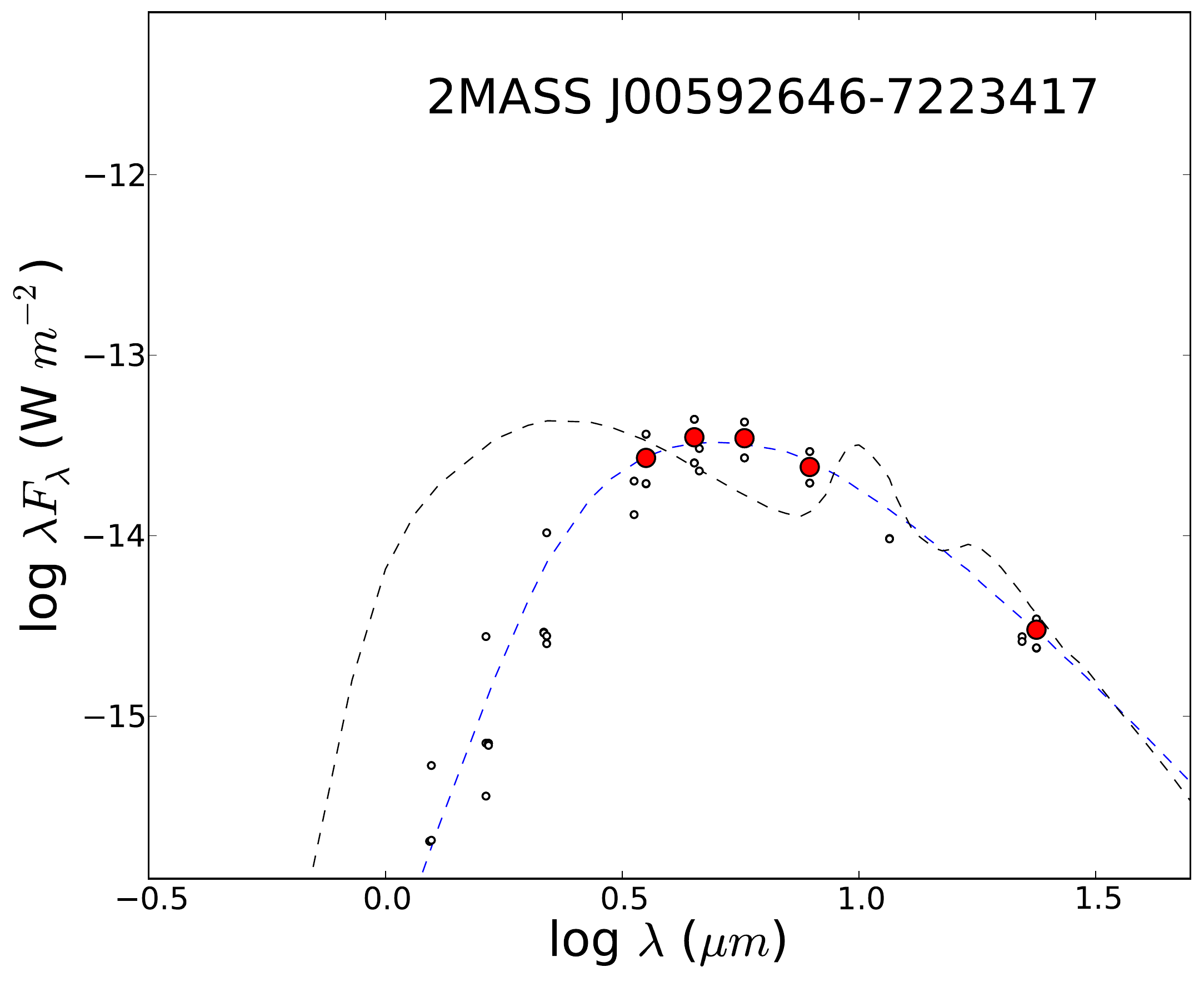}}
 \subfloat[]{\includegraphics[width=0.49\textwidth,height=0.49\textheight,keepaspectratio]{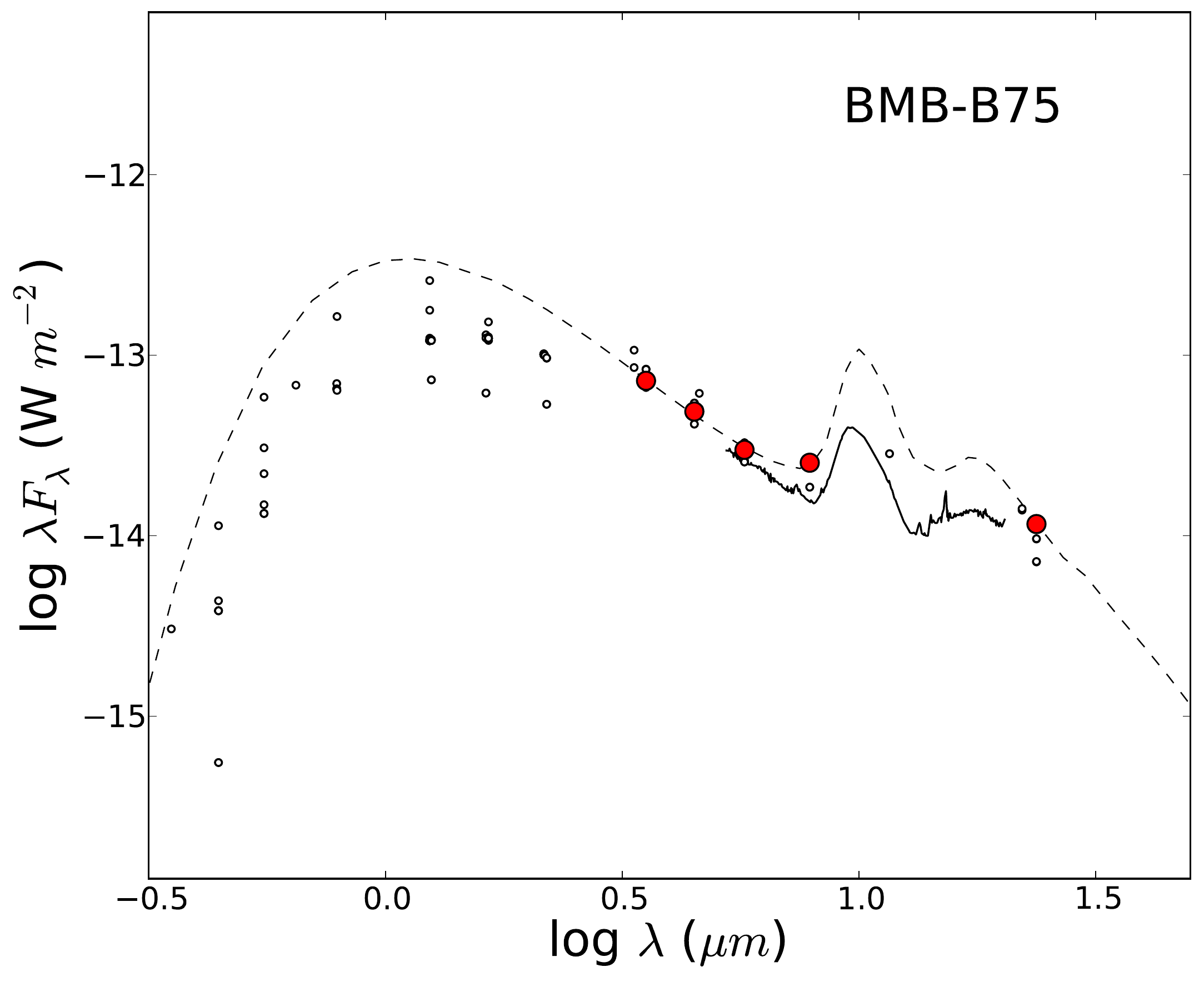}}\\ \vspace{-0.5cm}
 \subfloat[]{\includegraphics[width=0.49\textwidth,height=0.49\textheight,keepaspectratio]{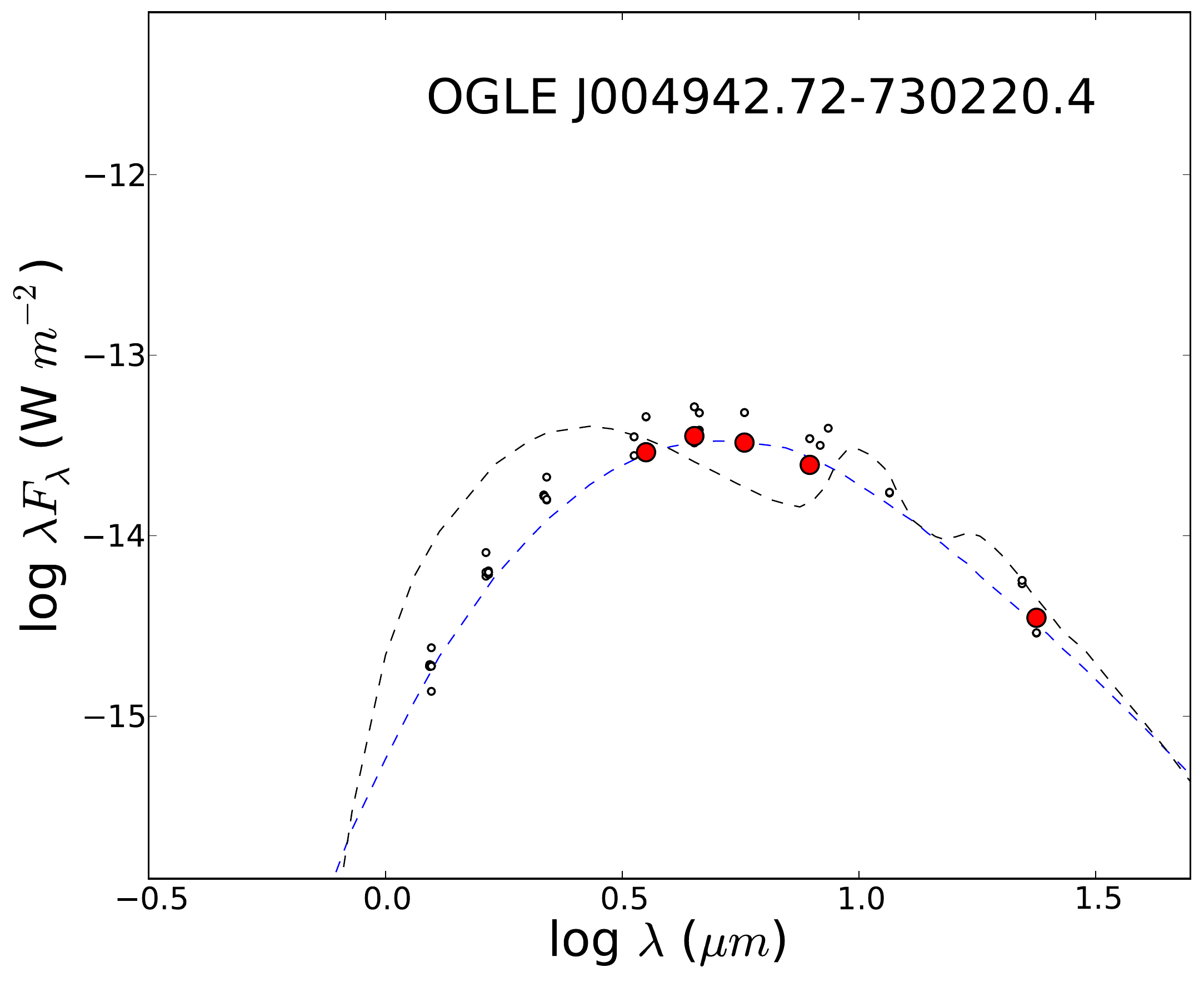}} 
 \subfloat[]{\includegraphics[width=0.49\textwidth,height=0.49\textheight,keepaspectratio]{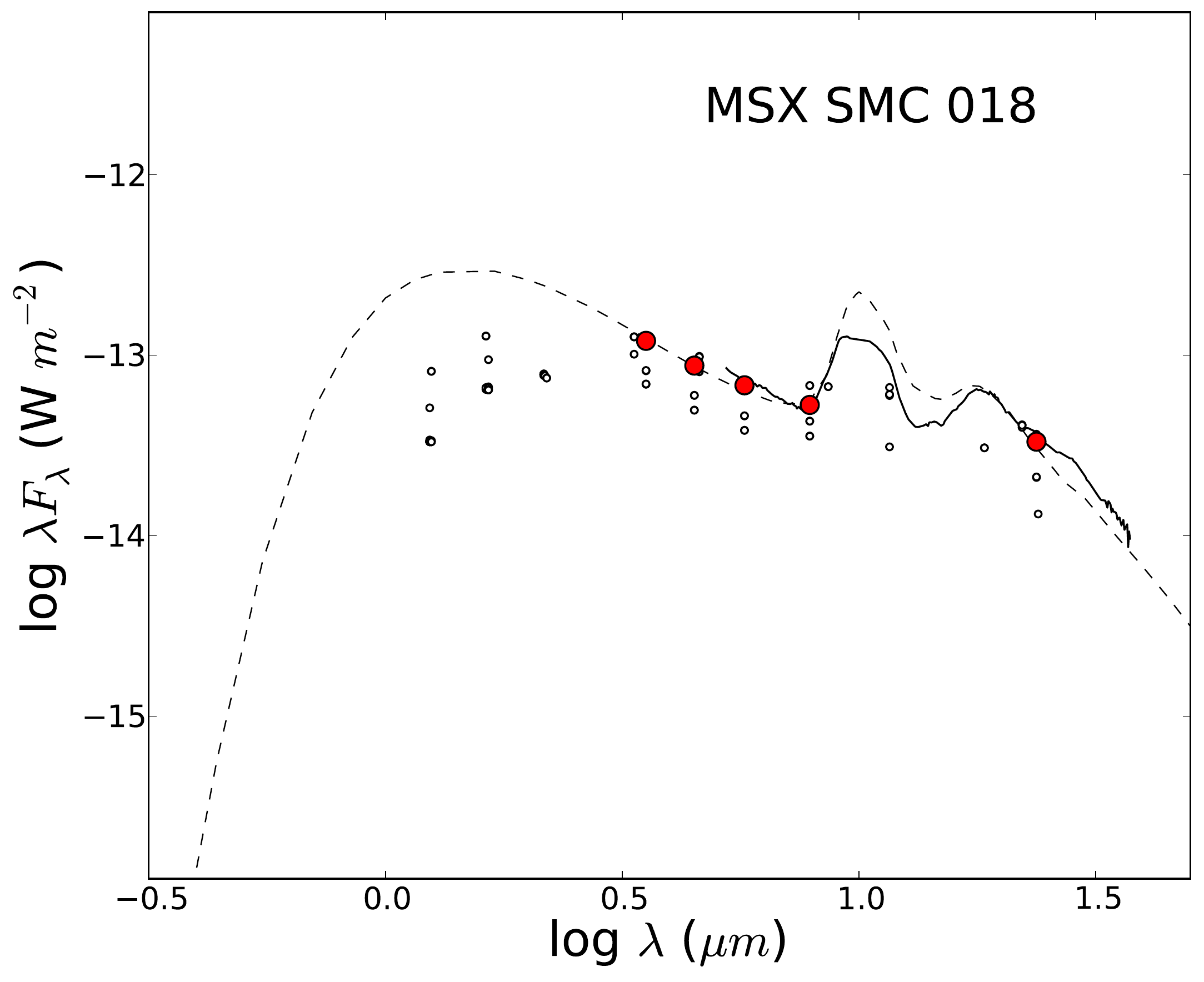}} 
 \caption{continued}
 \end{figure*}

\subsubsection{OH column density}

The minimum OH column density required to pump the 1612-MHz OH maser is around 1.3 $\times$ $10^{19}$ cm$^{-2}$ \citep{1976ApJ...205..384E}. Using our derived mass-loss rate (discussed in Section 4.5) and making several assumptions, we have calculated the expected OH column density of IRAS 00483$-$7347. Assuming the outflowing material is exclusively H$_2$ in an r$^{-2}$ distribution and using the inner OH shell radius determined for NML Cyg and VY CMa of 3.0 $\times$ 10$^{15}$ cm \citep{1979ApJ...233..119B}, this yields a value for the H$_2$ column density. While the radius of NML Cyg has been confirmed using mainline OH maser emission \citep{2004MNRAS.348...34E}, this value has been found to be larger ($\sim$ 300 au) for the more unique source VY CMa \citep{1982ApJ...253..199B}. NML Cyg and VY CMa have derived luminosities and mass-loss rates similar to those of IRAS 00483$-$7347 listed in Section 4.5 \citep{2011A&A...526A.156M,2016AJ....151...51S}, and should have a similar inner OH shell radius. Assuming a ratio of 5000:1 for the ratio of H$_2$ to OH \citep{2005IAUS..231..499O}, and scaling the abundance by 1/5 for metallicity, the fastest expansion velocity that would yield a value above the minimum column density for OH maser emission is 2.8 km s$^{-1}$. While slow for solar metallicity sources, it is still unclear if this value is realistic for the SMC's lower metallicity sample. From the \citet{2017MNRAS.465..403G} expansion velocity relation, the expected value, given the source's luminosity and metallicity is 5.4 km s$^{-1}$. This may suggest that the column density of the source has dropped below the masing threshold. The ratio of OH-to-H$_2$ remains the largest uncertainty in this calculation, as \citet{2005IAUS..231..499O} describes estimates more as order-of-magnitude approximations. Whether the abundance also scales linearly with the metallicity is also unclear. Regardless, we are observing sources very near the threshold for maser emission. The determining factor in whether maser emission is seen in the Galactic and LMC maser samples is the supply of IR photons to the system. Within the SMC, it seems that this becomes a two-body problem, where the maser strength drops with the metallicity squared, from a dependence on both the dust emission and the lower abundance of OH with respect to H$_2$.

\subsection{Mass-Loss and evolutionary phase}

It is expected that metal-poor AGB stars need longer to evolve towards cool temperatures and strong pulsations \citep{1991ApJ...375L..53B}. This would result in more massive white dwarfs as the core has more time to grow. Yet, observations have shown that the white dwarf mass functions between the SMC, LMC, and the Galaxy are similar \citep{2004ApJ...614..716V,2007ApJ...656..831V,2007ASPC..372...85W}. This would mean that the onset and strength of the dust-driven wind phase is similar within the Galaxy and the Magellanic Clouds \citep{vanLoon:2008iv}. However this is only the case for lower-mass AGB stars. The white dwarf mass function does not constrain the strength or duration of the superwind phase in either massive AGB stars or RSGs, which dominate our OH/IR samples in the Magellanic Clouds. With a shorter superwind phase, we would expect fewer sources exhibiting circumstellar maser emission.

 \begin{figure*}
 \centering
 \includegraphics[width=0.49\textwidth]{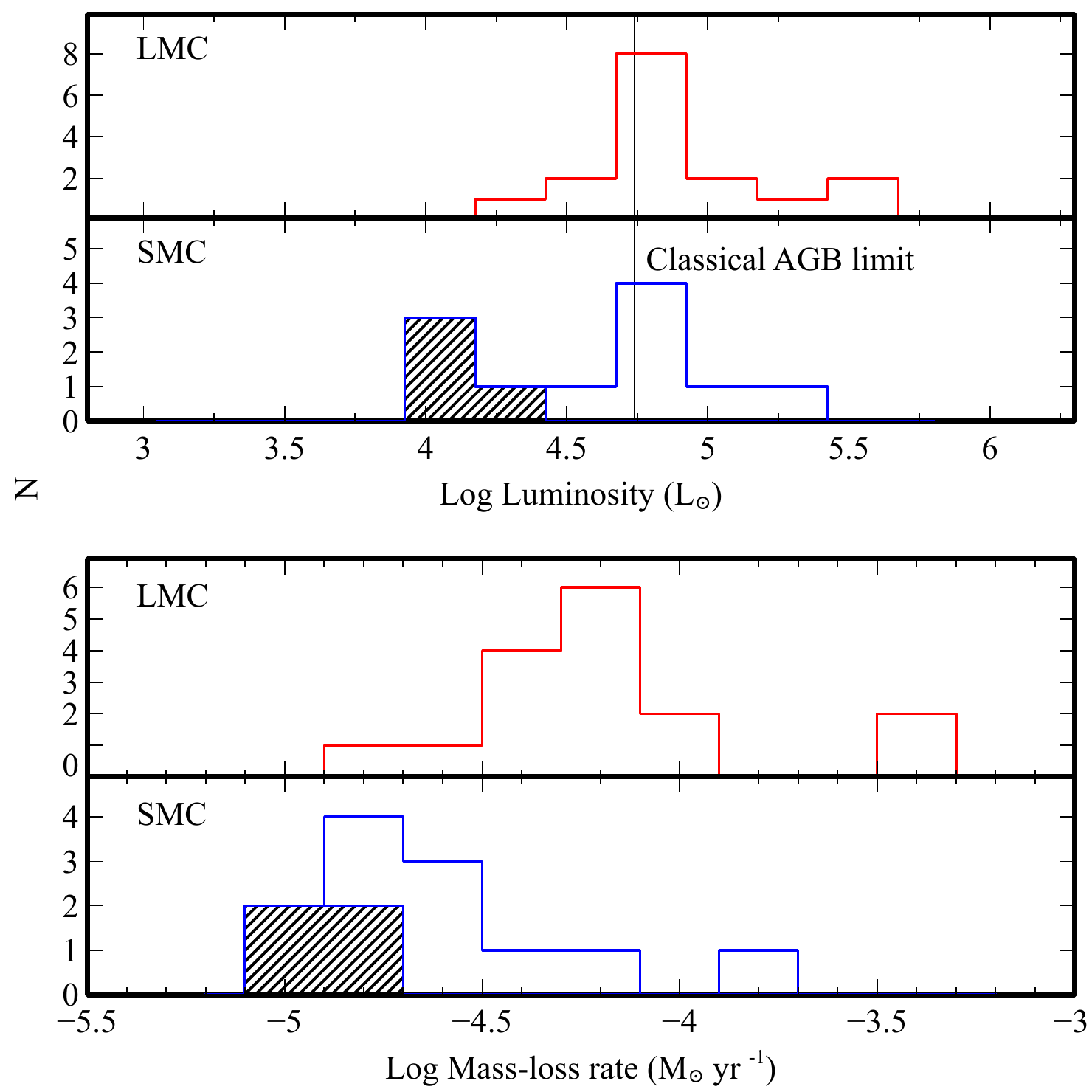}
 \includegraphics[width=0.49\textwidth]{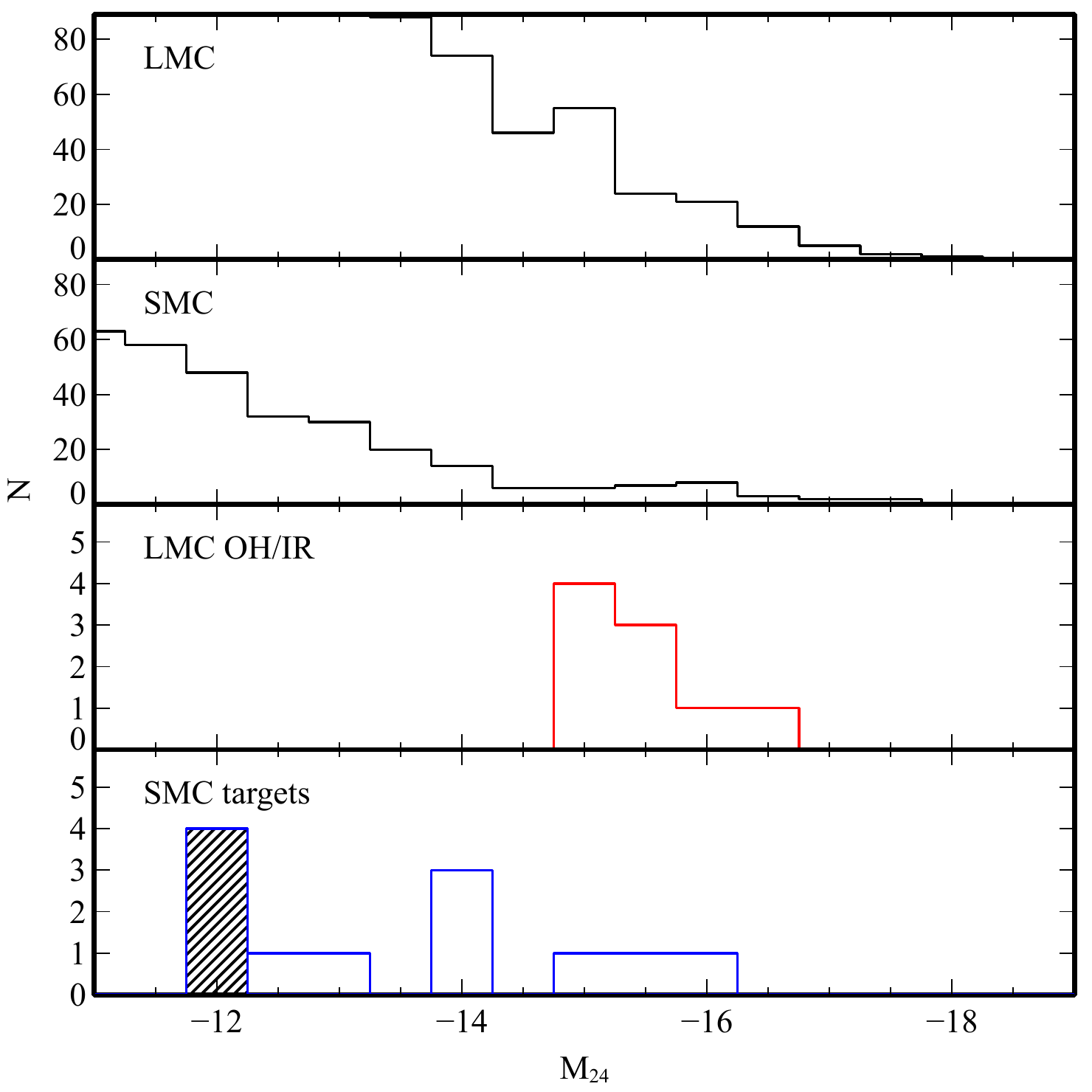}
 \caption[Luminosity distributions of our stellar samples]{\textit{Left}: Luminosity and mass-loss rate distributions of our SMC sample, and LMC OH/IR stars. \textit{Right}: The absolute 24 $\mu$m flux distribution of the SAGE-LMC and SAGE-SMC samples along with those of the LMC OH/IR and SMC samples. In both figures, our suspected carbon stars are designated with hatched lines.}
 \label{lum_hist}
\end{figure*}

Using the same SED fitting method as we applied to our LMC sample (see Goldman et al. 2017), we have fit the SEDs of our SMC sources (Fig. \ref{smc_seds}) and calculated luminosities and mass-loss rates for our SMC sample assuming a gas-to-dust ratio of 1000 \citep{2008AJ....136..919B,2004A&A...423..567B,2009ApJ...690L..76G}. The results show mass-loss rates $\sim 10^{-5}$ M\textsubscript{$\odot$} yr$^{-1}$, and luminosities typically $\sim 50,000$ L$_{\odot}$, similar to the LMC sample (Table \ref{table_smc_sed_results}). Due to the assumptions within the modeling of spherical symmetry, distance, optical properties, zero drift speed, and the fact that the model is an idealised system, it would be misleading to quote formal errors for these values. The luminosities are mostly affected by geometry and variability, but are generally accurate to within a factor two. The mass-loss rates are much less certain, and carry a systematic uncertainty of an order of magnitude due to our poor knowledge of the optical properties of the circumstellar grains \citep{2011A&A...532A..54S,Srinivasan:2016jp}. The table also shows expansion velocities calculated using the relation from \citet{2017MNRAS.465..403G}. While this relation has only been tested in metallicity environments between one half and twice solar, we expect the relation to give a good approximation within this metallicity regime. It is possible that at lower metallicity, a different set of dust optical constants may be better suited to reproduce the observed silicate features. However, implementing these changes would compromise the integrity of the comparison of samples, and thus we have used the same set of \textsc{dusty} models and dust optical constants as in the LMC analysis.

\begin{figure*}
 \centering
 \includegraphics[width=0.49\textwidth]{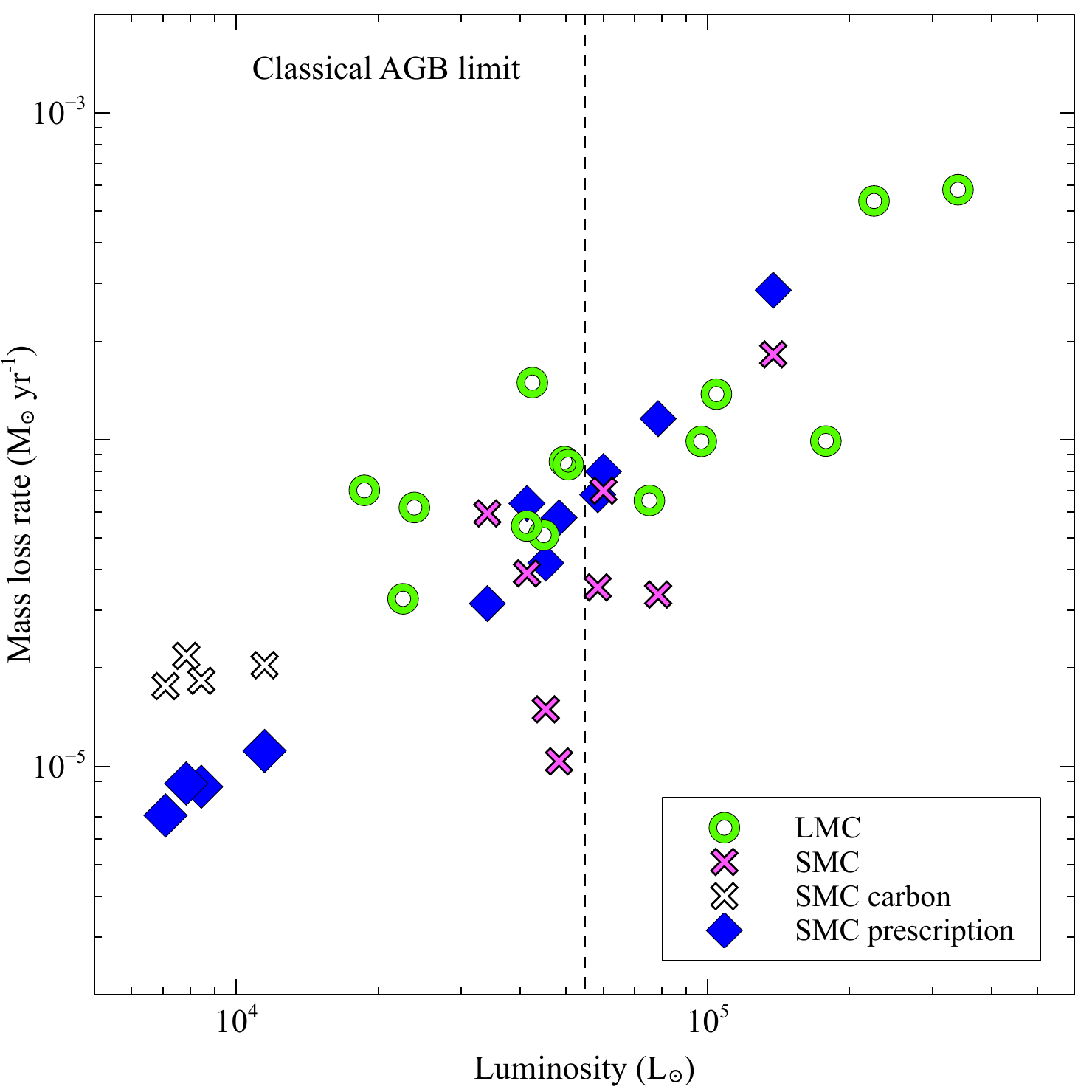}
 \includegraphics[width=0.49\textwidth]{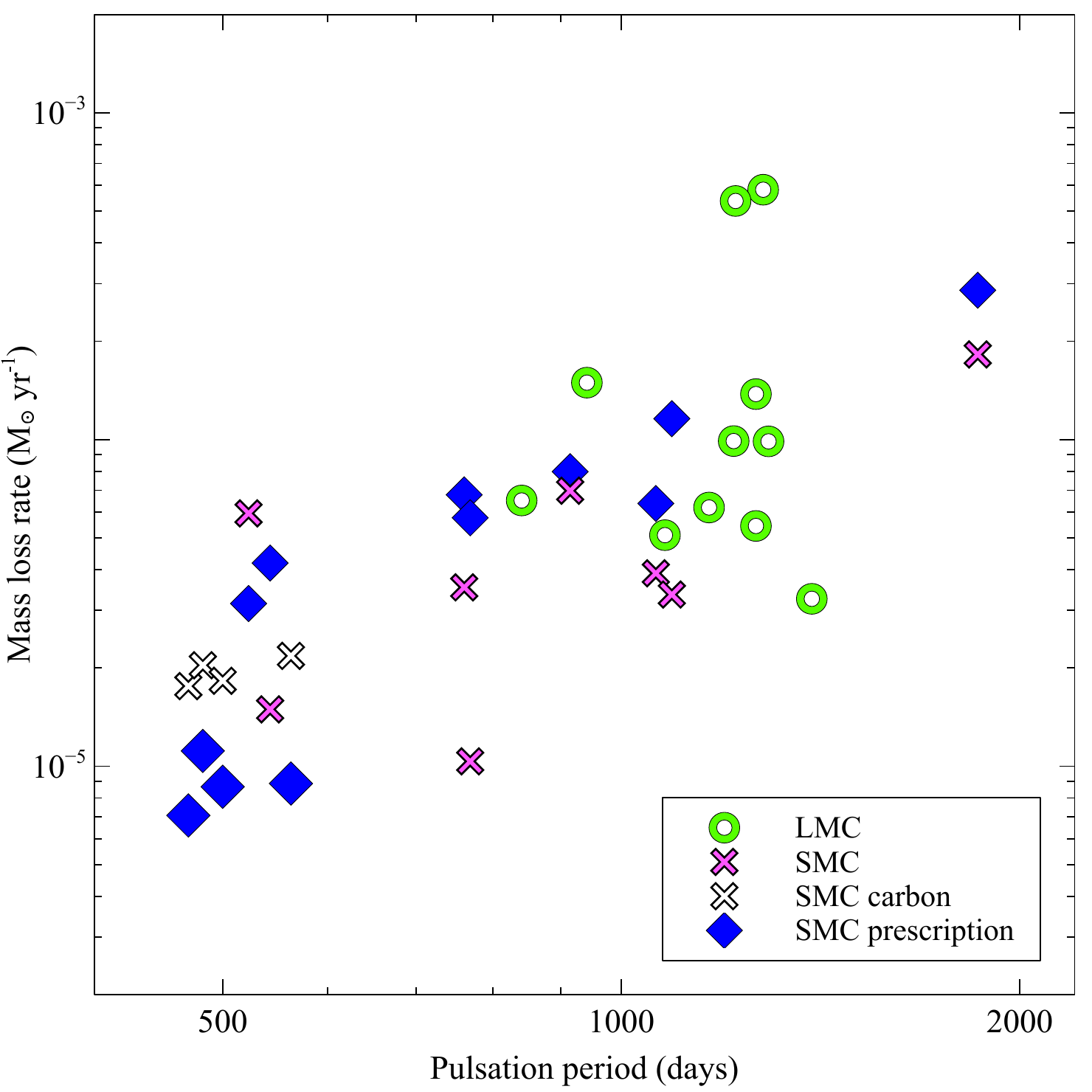}
 \caption[The mass-loss rate as a function of luminosity and pulsation period in the SMC]{The mass-loss rate as a function of luminosity (\textit{Left}) and pulsation period (\textit{Right}) for the LMC OH/IR sample and SMC sources. Also shown are the mass-loss rates calculated for the SMC sample using the \citet{2017MNRAS.465..403G} mass loss prescription (SMC prescription). A pulsation period of 500 d is assumed for the calculation of the prescription mass-loss rate for 2MASS J00592646-7223417. }
 \label{smc_M_vs_L} 
\end{figure*}

While there are only three sources (IRAS 00483$-$7347, IRAS 01074$-$7140, and MSX SMC 018) with comparable absolute flux densities at 24 $\mu$m to the LMC sample, there are several sources with comparable luminosities (Fig. \ref{lum_hist}). Our most luminous source IRAS 00483$-$7347 has a long pulsation period, late spectral type, a large Rb enhancement, typical of an AGB star that has undergone HBB. Yet the source has a high luminosity of 138,000 L$_{\odot}$ suggestive of an RSG, which is more likely to indicate that the source is likely a super-AGB star.

When comparing the mass-loss rates and luminosities for the SMC sample to the LMC OH/IR sample (Fig. \ref{lum_hist}, \ref{smc_M_vs_L}), we see that the higher luminosity SMC sources ($ > 2\times 10^{4}$ L$_{\odot}$) tend to have a lower mass-loss rate at a given luminosity (except for IRAS 00483$-$7347). This is not surprising given the SMC sources are not as optically thick as the LMC sources. We have also included the SMC mass-loss rates calculated using the \citet{2017MNRAS.465..403G} mass loss prescription (SMC prescription). Using the prescription we assume a gas-to-dust ratio of 1000, yet this parameter does not significantly affect the calculated mass-loss rate. The values we calculate are typically higher than what we calculate from SED fitting, except for several sources that have SEDs that do not conform well to any of the models within our grids (2MASS J01033142$-$7214072, OGLE SMC-LPV-14322, 2MASS J00592646$-$7223417, OGLE J004942.72$-$730220.4). We have modelled these four sources using a similar grid of \textsc{dusty} models that use the carbon dust grains from \citet{1988ioch.rept...22H}. We have over-layed the best fit carbon-SED for these sources and find that they better reproduce the observed SED. This along with their position with respect to the \textsc{parsec-colibri} isochrones (Fig. \ref{smc_L_vs_J-K}) and their low luminosities may signify that they are in fact less evolved non-maser-emitting carbon stars. Within the [3.6]$-$[8.0] vs. absolute flux density at 24 $\mu$m CMD (Fig. \ref{in_SMC}), these less luminous sources occupy an area typical of carbon stars, providing further evidence for this conclusion. 

Looking at mass-loss rates versus pulsation period (Fig. \ref{smc_M_vs_L}) we see a spread of values for the SMC that span our other samples. We also see an increase in the mass-loss rate with pulsation period for the SMC sample. There are several sources with comparable luminosities, pulsation periods and mass-loss rates, suggesting that the SMC sources are in a similar evolutionary stage within the superwind phase. An example being the SMC source BMB-B75 and the LMC source IRAS 05402$-$6956, with luminosities of 58,000 and 51,000 L$_{\odot}$, pulsation periods of 1453 and 1393 d, and mass-loss rates of 3.5 and 8.4 $\times$ 10$^{-5}$ M$_{\odot}$ yr$^{-1}$, respectively. 

From our new relation for the circumstellar expansion velocity \citep{2017MNRAS.465..403G}, we have calculated the expected  expansion velocities of our SMC sample (shown in Table \ref{table_smc_sed_results}) assuming a metallicity of one fifth solar. Typical expansion velocities for these sources are expected around 3 km s$^{-1}$, yet it is possible that as the strength of radiation pressure wanes, other wind-driving forces take over, resulting in underestimated predictions for the wind speed. We have also modelled SMC sources with \textsc{dusty} models calculated with a simple inverse square density distribution (as opposed to our SEDs calculated using a full hydrodynamical solution) and found negligible difference between the results. This may indicate that the material is blown out without the need for pulsation-enhancement. With a slower wind we also expect more efficient stimulated emission as a result of higher density. This higher rate of collisions may then make it easier to quench the maser. Another scenario worth mentioning is the possibility of binarity in the SMC sample. While we have no evidence to suggest this, symbiotic Miras tend to have weaker maser emission as a result of the decreased coherence within the OH shell \citep{1995MNRAS.276..867S}. This may provide another possible explanation for a decrease in masing sources within the SMC. We can thus conclude that there is no strong evidence for AGB stars and RSGs in the metal-poor SMC, at $\sim0.2$ Z$_\odot$, to experience weaker mass loss than those in the more metal-rich systems, the LMC ($\sim0.5$ Z$_\odot$) and Galactic Bulge and Centre ($\sim0.5$--2 Z$_\odot$).

\subsection{Future prospects}

The Galactic Australian Square Kilometre Array Pathfinder (GASKAP) Spectral Line Survey is an upcoming high spectral resolution (0.2 km s$^{-1}$) ASKAP survey of the 21-cm HI, and 18-cm OH lines, in the Milky Way galactic plane, the LMC, the SMC, and the Magellanic Bridge and Stream \citep{2013PASA...30....3D}. GASKAP will survey these regions with sensitivities better than that of Parkes (flux density sensitivity around 0.5 mJy with a 200 hour integration). As the entire SMC will fit within the field of view of ASKAP, this is feasible and there is therefore a good chance that GASKAP will detect OH maser emission from the most luminous candidates wherever they reside in the SMC. It was predicted by \citet{2015aska.confE.125E}, based on the relation between 1612 MHz OH maser emission and mass-loss rate \citep{1996MNRAS.279...32Z}, that around two dozen OH maser sources should be detectable in the SMC at a 3-$\sigma$ sensitivity of 0.1 mJy. Following the same approach, for the range in mass loss parameters listed in Table 4 for our SMC sample we would predict $3-29$ mJy. We are therefore somewhat more optimistic about the prospects for GASKAP to detect more than a few OH masers in the SMC, or for it to confirm that the maser mechanism breaks down at the metallicity of the SMC. This survey will also be a precursor to much larger surveys using the Square Kilometre Array, which will reveal the full extent of the maser samples in these regions or show that the maser mechanism breaks down at the metallicity of the SMC.

\section{Conclusions}

Using the properties of past OH maser host stars, we have identified and observed the brightest SMC sources likely to exhibit circumstellar maser emission. Our new, deep 1612-MHz OH maser observations with single and interferometric instruments have not resulted in any clear maser emission from our targeted or in-field sources, yet a number of maser-emitting sources may still lie below our observation detection thresholds. Assuming the masing mechanism is the same within the Galaxy and Magellanic Clouds, the top two candidates for maser emission have upper limits for maser efficiency (converting $F_{35}$ to $F_{\textrm{OH}}$) of 3.6 and 4.3\%. These values are dramatically lower than the typical maser efficiency around 23\%. This may point to differences within the circumstellar environments of evolved stars at lower metallicity that may include a low OH abundance or strong interstellar radiation. We cannot adequately quantify the effects of metallicity on the strength of circumstellar OH maser emission. We suspect that some of these sources are near the OH column density threshold for maser emission and also suspect that maser emission may still lie below our detection threshold. It is also possible that the OH masing phase does not last as long in the SMC as at higher metallicity, at least for massive AGB and RSGs. This could have implications for the contribution from these stars to the total dust budget in the galaxy, and for the contribution in which the RSGs undergo core collapse. Future observation with the GASKAP survey \citep{2013PASA...30....3D} will be able to add clarity to our understanding of circumstellar maser emission and the effects of metallicity. 

\section*{Acknowledgments} 
We would like to thank the referee, Dieter Engels, for his helpful and constructive comments. We would like to thank the support staff at CSIRO for their help with the observations and Michele Costa for his help with the bootstrapping code, and Leen Decin for her helpful discussion on circumstellar environments. And we would also like to thank Phillip Edwards for generously allocating directors discretionary time to the project. AN and JvL acknowledge an award from the Sir John Mason Academic Trust in support of theoretical investigations of dusty winds and evolved star populations. JvL, JO and HI acknowledge awards from the Daiwa Foundation, 11595/12380, and the Great Britain Sasakawa Foundation, 5219, in support of their population studies of astrophysical masers. PAW acknowledges a research grant from the South African National Research Foundation (NRF). AN acknowledges the support from the project STARKEY funded by the ERC Consolidator Grant, G.A. No. 615604. JFG is supported by MINECO (Spain) grant AYA2014-57369-C3-3 (co-funded by FEDER). This paper makes use of the SIMBAD database of the CDS.

\bibliographystyle{mn2e_new}
\bibliography{references}

\onecolumn
\appendix
\renewcommand{\thefigure}{A\arabic{figure}}
\setcounter{figure}{0}

\begin{table*}
\begin{flushleft}
\section{Past SMC maser targets}
There have been two previous 1612-MHz maser surveys in the SMC conducted by \citet{2004MNRAS.355.1348M} and \citet{1992ApJ...397..552W}. The observations by \citet{2004MNRAS.355.1348M} targeted two sources, IRAS 00483$-$7347, which we have re-observed with a higher sensitivity, and IRAS 00591$-$7307 which we have reprocessed to present a complete and uniform dataset. The observations by \citet{1992ApJ...397..552W} focused on 15 sources targeted for having high IRAS 25 $\mu$m flux densities and that fit the criteria: $0.0 < $ log $ (F_{25}/F_{12}) < 0.5$. These observations achieved noise levels typically $\sim$ 40 mJy and did not result in any clear maser detections in the SMC. The 15 targets are composed of an evolved AGB star, a Seyfert 2 Galaxy, a star forming region, 2 foreground stars, 2 YSOs, 3 H\,{\sc ii} regions, and 5 sources whose identity is still unclear (Table A1). We have presented new observations for the re-observed AGB star, HV 12149. The foreground sources are the red giant IRAS 00542$-$7334 (CM Tuc) \citep{1983A&AS...53..255P,2015A&A...578A...3G}, and the F star IRAS 00490$-$7125 (HD 5028) \citep{2006AstL...32..759G}. The unknown source IRAS 00435$-$7339 lies near a star-forming region. The Seyfert 2 galaxy that we observed is the well studied peculiar source IRAS 00521$-$7054 \citep{2010A&A...518A..10V,2014ApJ...795..147R}. None of the unknown sources, or the source IRAS 00477$-$7322 have 2MASS counterparts.

\end{flushleft}
\centering
\caption[wood]{Targets from the 1612-MHz OH maser survey by \citet{1992ApJ...397..552W}.} 
\begin{tabular}{ l c c}
\hline
Target & Type & Reference \\
\hline
HV 12149 &
AGB (SpT=M8.5) &
{\citet{2015A&A...578A...3G}} \\
IRAS 00430$-$7326 &
YSO &
\citet{2010AJ....139.1553V} \\
IRAS 00435$-$7339 &
? &
\\
IRAS 00466$-$7322 &
H\,{\sc ii} region& 
\citet{2012ApJ...755...40P} \\
IRAS 00477$-$7322 &
? &
\\
IRAS 00477$-$7343 &
H\,{\sc ii} region &
\citet{2004AJ....128.2206I} \\
IRAS 00490$-$7125 &
F-type star &
\citet{2006AJ....132..161G} \\
IRAS 00517$-$7240 &
? &
\\
IRAS 00521$-$7054 &
Seyfert 2 galaxy&
\citet{1987NASCP2466..241E}\\
IRAS 00542$-$7334 &
Foreground AGB &
{\citet{2015A&A...578A...3G}} \\
IRAS 00553$-$7317 &
? &
\\
IRAS 01039$-$7305 &
YSO &
\citet{vanLoon:2008iv} \\
IRAS 01126$-$7332 &
Star-forming region &
\citet{2013MNRAS.432.1382B} \\
IRAS 01208$-$7532 &
? &
\\
IRAS 01228$-$7324 &
H\,{\sc ii} region & 
{\citet{1985A&A...145..170T}}\\
\hline
\end{tabular}

\vspace{-0.5cm}

\begin{flushleft}
\section{YSO OH maser spectra}
After observing, it was found that two sources within our maser survey fields were in fact YSOs. These sources were suggested as YSO candidates by \citet{vanLoon:2008iv} and then confirmed by \citet{2013MNRAS.428.3001O}. We present the 1612-MHz spectra for the sources IRAS 01039$-$7305 and IRAS 01042$-$7215. They were observed for 7.4 and 6.3 hours, respectively. YSO's are expected to have their strongest emission in the mainlines transitions, but will also exhibit maser emission at 1612 MHz. We do not see any clear maser emission in either source but do see several large narrow spikes, likely a result of RFI. 
\end{flushleft}
\vspace{-0.25cm}
\end{table*}
\begin{figure}
 \centering
 \includegraphics[height=8 cm]{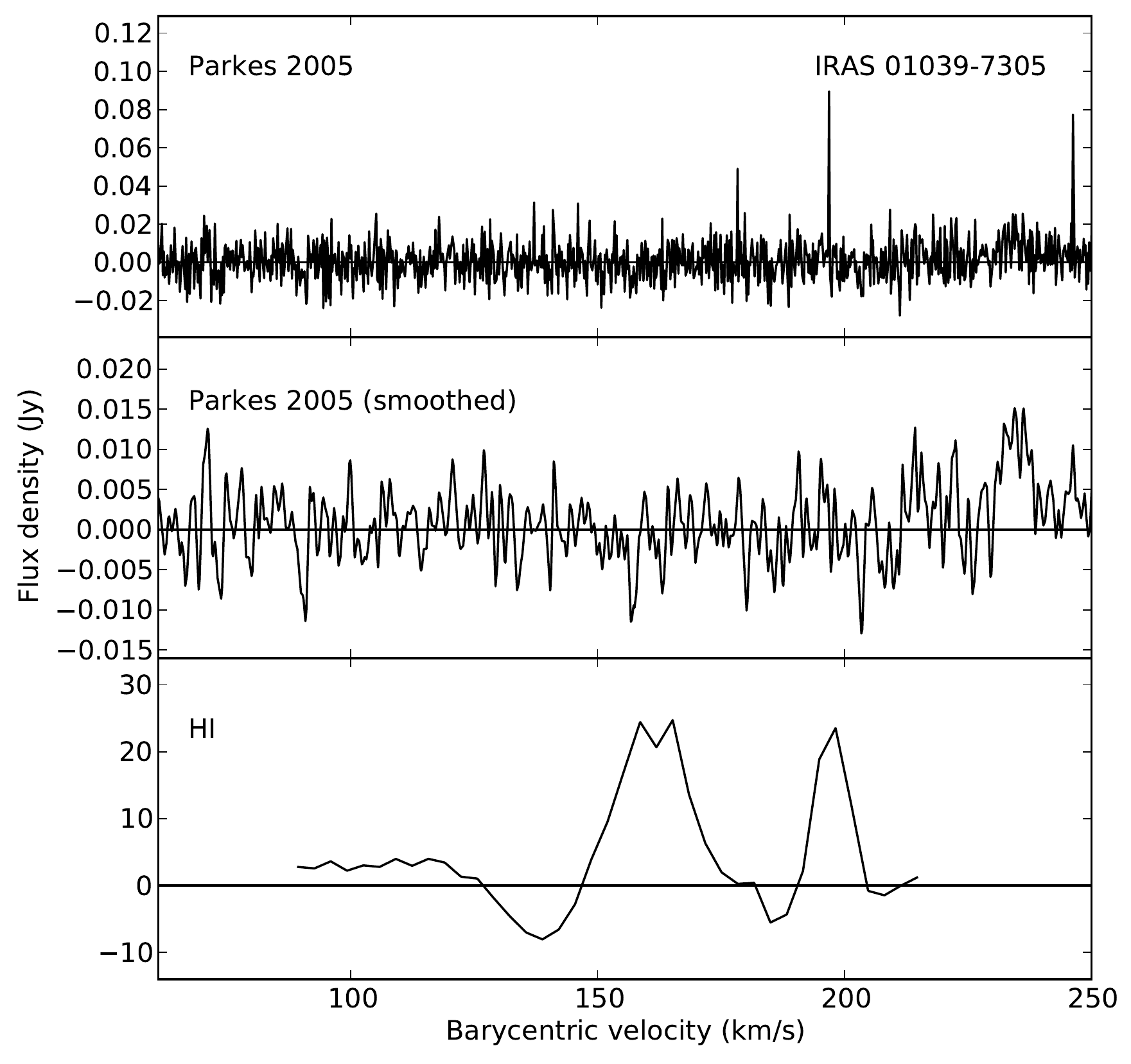}
 \includegraphics[height=8 cm]{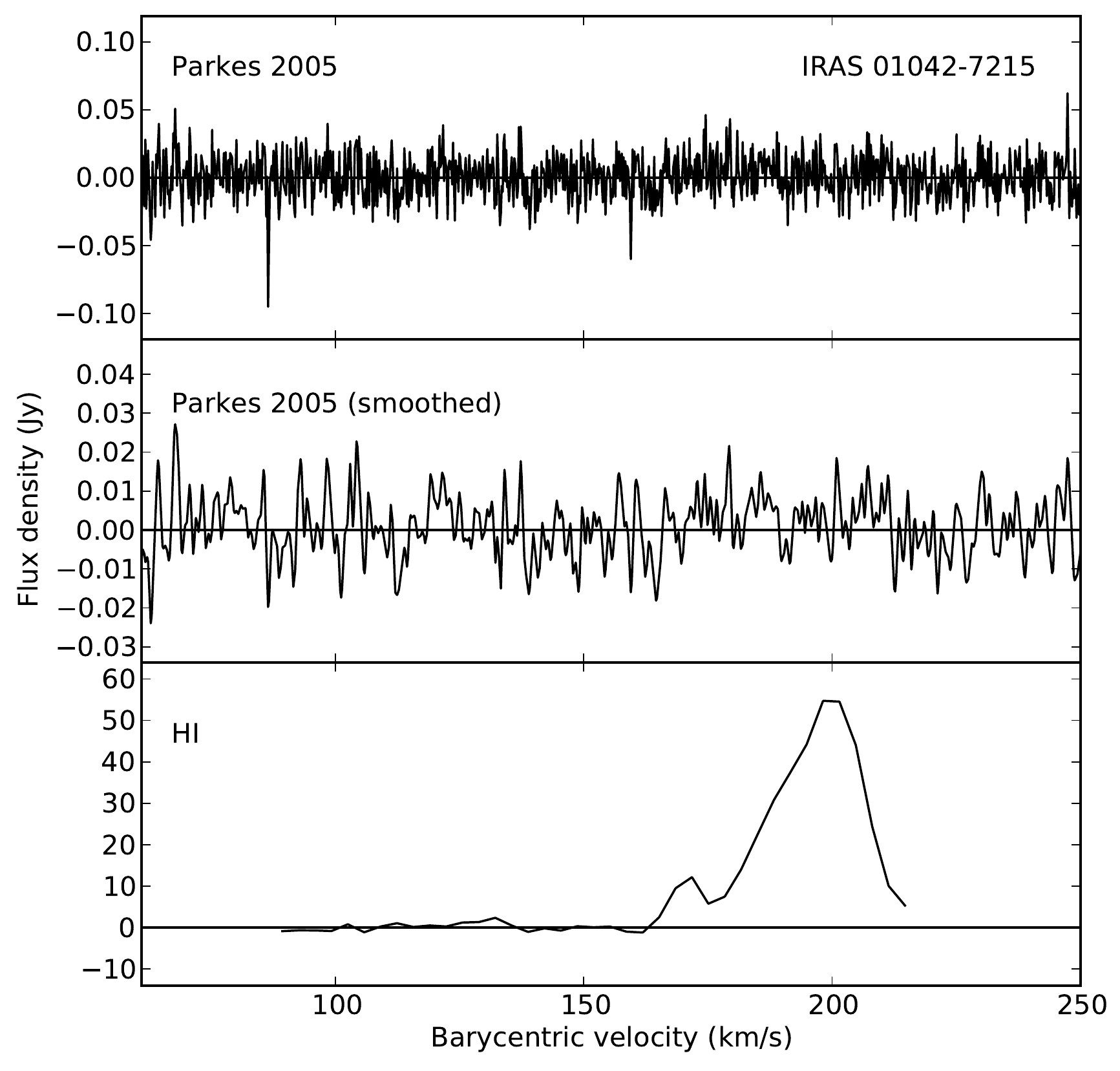}
 \caption{OH 1612 MHz maser observation of IRAS 01039$-$7305 and IRAS 01042$-$7215. As in Figure 2, the \textsc{HI} spectra are presented for velocity reference.}
 \label{yso spectra}
\end{figure}

\clearpage 
\renewcommand{\thefigure}{C\arabic{figure}}
\setcounter{figure}{0}

\begin{figure}
 \begin{flushleft}
 \section{Mainline ATCA observations}
Each of our ATCA observations observed multiple OH maser lines. In addition to the 1612-MHz OH satellite line, the observations also targeted the 1665- and 1667-MHz OH mainlines, which probe regions closer to the star, within the 1612-MHz maser profile. The 1612-MHz maser dominates at longer path lengths \citep{1991MNRAS.252...30G}, where the 1665-MHz mainline emission can come from the OH being located in a high density region (close to the star). High densities ($\gtrsim 10^7$ cm$^{-3}$) suppress the 1667-MHz emission compared to the 1665-MHz emission. The 1665-MHz spectra for each of the sources observed with the ATCA have been resampled to a velocity resolution of 0.5 km s$^{-1}$ and are displayed below the 1612-MHz spectra in the figure below. The much more sensitive Parkes observation of IRAS 00483$-$7347 is shown for the 1612-MHz spectrum. 

\vspace{0.3cm}
 \label{lastpage}
 \end{flushleft}
 \centering
 \includegraphics[width=0.99\textwidth,height=0.99\textheight,keepaspectratio]{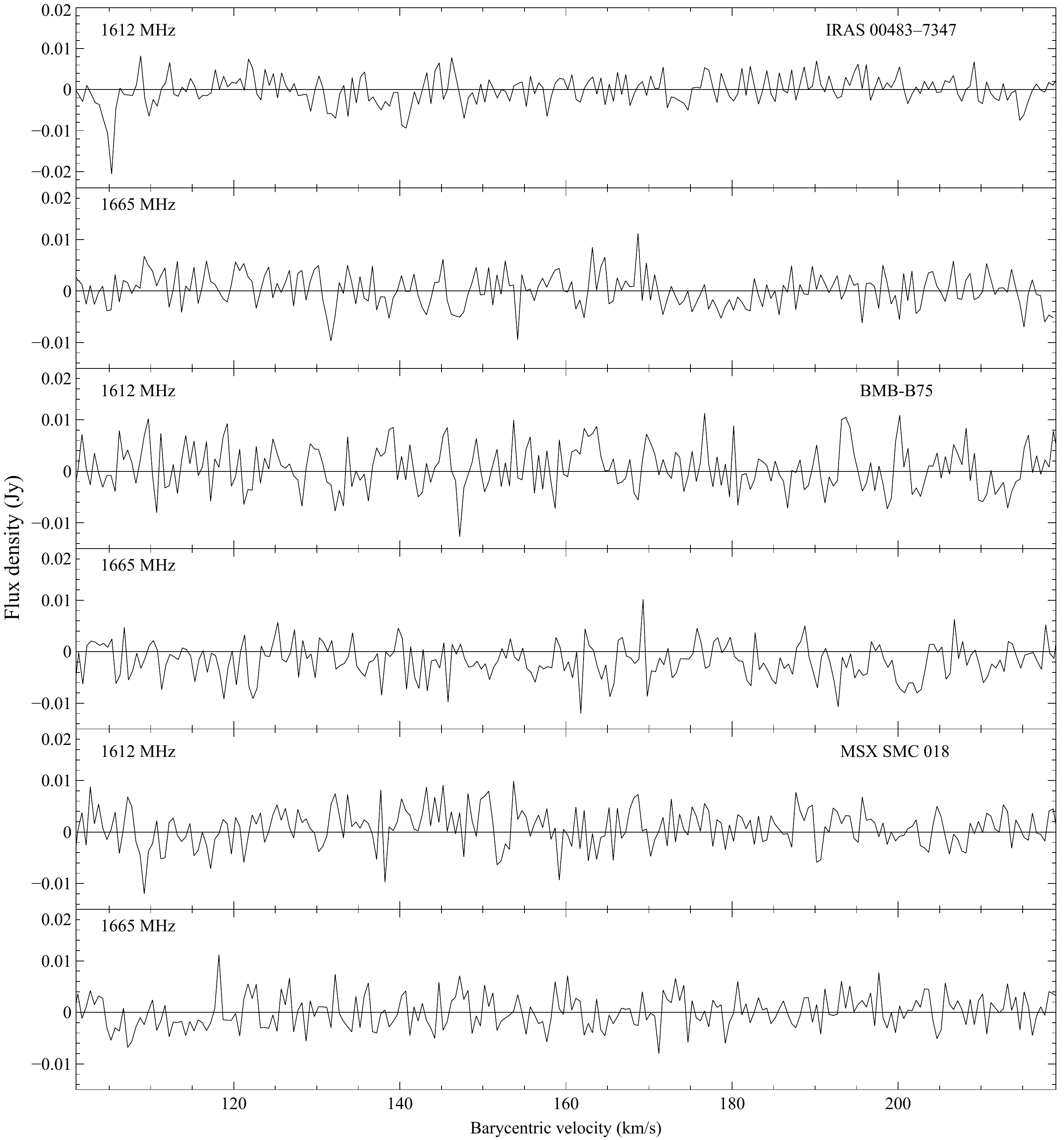}
 \caption{The 1612- and 1665-MHz OH maser spectra of SMC sources observed with the ATCA. }
 \label{mainlines}
\end{figure}

\end{document}